\documentclass[fleqn,usenatbib]{mnras}

\usepackage{newtxtext,newtxmath}
\usepackage[T1]{fontenc}
\usepackage{ae,aecompl}
\usepackage[utf8x]{inputenc}
\usepackage{hyperref}
\usepackage{bbm}
\usepackage{graphicx}	
\usepackage{amsmath}	
\usepackage{amssymb}	

\newcommand{\m}{m_{1}}
\newcommand{\mm}{m_{2}}
\newcommand{\mmm}{m_{3}}
\newcommand{\R}{R_{1}}
\newcommand{\RR}{R_{2}}
\newcommand{\ProtA}{P_{\mathrm{rot,A}}}
\newcommand{\ProtB}{P_{\mathrm{rot,B}}}
\newcommand{\tvA}{t_{\mathrm{v,A}}}
\newcommand{\tvB}{t_{\mathrm{v,B}}}

\newcommand{\kappaA}{\kappa_{\mathrm{A}}}
\newcommand{\kappaB}{\kappa_{\mathrm{B}}}
\newcommand{\ain}{a_{1}}
\newcommand{\aout}{a_{2}}
\newcommand{\rin}{r_{1}}
\newcommand{\rout}{r_{2}}
\newcommand{\Hcal}{\mathcal{H}}
\newcommand{\secHcal}{\langle\langle \Hcal \rangle \rangle}
\newcommand{\vrin}{\mathbf{\vec{r}_{1}}}
\newcommand{\vrout}{\mathbf{\vec{r}_{2}}}
\newcommand{\itot}{i_{\mathrm{tot}}}
\newcommand{\Msun}{M_{\odot}}
\newcommand{\Rsun}{R_{\odot}}
\newcommand{\BDlangle}{\bigg \langle \bigg \langle} 
\newcommand{\BDrangle}{\bigg \rangle \bigg \rangle} 
\newcommand{\tahalf}{t_{a,1/2}}

\newcommand{\tfA}{t_{Fs}}
\newcommand{\tfB}{t_{Fp}}
\newcommand{\OmAx}{\Omega_{s,x}}
\newcommand{\OmAy}{\Omega_{s,y}}
\newcommand{\OmAz}{\Omega_{s,z}}
\newcommand{\OmBx}{\Omega_{p,x}}
\newcommand{\OmBy}{\Omega_{p,y}}
\newcommand{\OmBz}{\Omega_{p,z}}
\newcommand{\Gcal}{\mathcal{G}}

\renewcommand{\i}{I}

\title[Dynamics of planets in binaries]{Revisiting the dynamics of planets in binaries: evolutionary timescales and the effect of early stellar evolution}

\author[Bayron Portilla-Revelo \& Jorge I. Zuluaga]{
Bayron Portilla-Revelo$^{1}$\thanks{E-mail: bayron.portilla@udea.edu.co} and 
Jorge I. Zuluaga$^{1}$\thanks{E-mail: jorge.zuluaga@udea.edu.co}
\\
$^{1}$Group for Computational Physics and Astrophysics (FACom) and Solar, Earth and Planetary Physics Group (SEAP)\\
Instituto de Física-FCEN, Universidad de Antioquia\\ 
Calle 70 No. 52-21, Medellín, Colombia.
}

\date{Accepted XXX. Received YYY; in original form ZZZ}

\pubyear{2018}

\begin{document}
\label{firstpage}
\pagerange{\pageref{firstpage}--\pageref{lastpage}}
\maketitle

\begin{abstract}
The discovery of planets in binaries is one the most interesting outcomes of planetary research. With the growing number of discoveries has also grown the interest on describing their formation, long-term evolution and potential habitability. In this work we revisit the dynamics of planets in S-type binary systems. For that purpose we develop explicit formulas for the secularized octupolar Hamiltonian, coupled with general relativistic corrections and non-conservative interactions. We implemented those formulas in an open-source package \texttt{SecDev3B}, that can be used to reproduce our results or test improved versions of the models. In order to test it, we study the long-term dynamical evolution of S-type binary planets during the pre-main-sequence phase of stellar evolution. During that phase, stellar radius significantly changes in timescales similar to secular timescales. We hypothesize that when close-encounters between the planet and its host star happens (e.g. via Lidov-Kozai effect), particularities in the secular formalism plus changes in stellar radius  may alter significantly the dynamical evolution. 
We study the well-known binary planet HD 80606b and found that an octupolar expansion of the conservative Hamiltonian is required to properly predict its dynamical evolution. We also apply the dynamical model, enriched with results coming from stellar evolutionary models, to demonstrate that in S-type systems around low-mass stars, with relative high inclinations ($\itot\ge 60^\circ$), moderate eccentricities ($0.2\le e\le 0.4$) and planets located around 1 AU, the evolution of stellar radius during the first few hundreds of Myr, alters significantly the timescales of dynamical evolution. 
\end{abstract}



\section{Introduction}
\label{sec:intro}

Planets in binaries are among the most interesting and unexpected discoveries in the era of exoplanets.  These triple (or multiple) gravitational systems are divided, according to their hierarchical configuration, into two categories \citep{Dvorak1986}: 1) {\em S-type systems}, where the planet orbits one member of the stellar binary ({\em S} stands for {\em satellite}) and the other star is an outer perturber; and 2) {\em P-type systems} in which both stars are closely orbiting each other with the planet moving on a wider orbit ({\em P} stands for {\em planet}).  

To the date of writing, 132 planets have been discovered around 91 binaries \citep{Schwarz2016}\footnote{See for instance the Catalogue of Exoplanets in Binary Stars Systems \url{https://www.univie.ac.at/adg/schwarz/multiple.html}}. Around 75\% of them correspond to S-type systems.

Planets in binaries are not only interesting because their dynamical properties (which we explore in this work) but also because they challenge our current understanding of planetary formation.  

In the case of isolated stars, the formation of planets is the result of two possibles mechanism: core accretion or disk instabilities \citep{Armitage2009}. The same mechanisms are in action in S-type systems, provided the separation of the outer companion is large enough to make its gravitational effect on the circumstellar disk, negligible. However, since we have discovered planets around compact S-type systems, namely, binaries with separations as small as $20-25$ AU (\citealt{Hatzes2003}; \citealt{Zucker2004}; \citealt{Mugrauer2005}), the outer companion is expected to have had a considerable effect on the coalescence of planetesimals.  Still, the existence of planets around this type of systems, evidences the fact that binarity does not preclude the subsequent formation of planets \citep{Haghighipour2010}.   

Independently of the mechanism that lead to their formation, planets in binaries appears to be more common than expected. The apparent lack of them in what seems to be {\em planet-less} binary systems, could be explained due to the biases of the discovering techniques \citep{Pelupessy2013}.

Numerous works have studied the dynamics of circumbinary systems, both using direct $N$-body calculations 
(see e.g. \citealt{Holman1999} and \citealt{Kostov2016}) and secular theories based on quadrupolar and octupolar order expansions of the perturbative Hamiltonian (see e.g. \citealt{Wu2003}; \citealt{Mardling2013}; \citealt{Naoz2013}; \citealt{Correia2016}). The secular approach has proved to be significantly more efficient than direct $N$-body methods for long integration times \citep{Naoz2016}. One of the major advantages of secular theories relies in the fact that relatively large time-steps (even greater than the orbital periods of the components) can be used in the computations.  This is possible thanks to the fact that rapidly varying quantities (e.g. mean anomalies) are properly removed from the Hamiltonian and hence from the equations of motion, by averaging-out over relevant time-scales. However, a major disadvantage of this approach is the need for the inclusions of octupolar order terms when both members of the inner binary are not equal in mass.


Mathematical expressions in this formalism are complex. 
This fact motivated us to develop a software package { \tt SecDev3B} that implements all those formulas, while adding new functionalities (see below).  This package will contribute to the community making our models and results (1) reproducible and (2) improvable. Its modular design and internal documentation will allow the future inclusion of new and better approximations.

But this paper is not only about the software, but also on what we discovered while developing and testing it. We report, for instance, a major difference in the evolution timescale of the well-known HD 80606b planet when the integration is made retaining the octupolar term of the secular Hamiltonian, in contraposition to those previously studied cases in which a quadrupolar expansion was used.

When integrations of hundreds of million years (Myr) are performed on this type of systems, there could be noticeable changes in the radius and internal properties of the stars and the planet.  This is particularly notorious during the pre and post main sequence evolution of the stellar components \citep{Carroll2007} and during the first hundreds of Myr following planetary accretion, especially in the case of giant planets \citep{Fortney2007}.

In the past, several works have explored the interplay between dynamics, stellar evolution and other aspects of stellar structure, especially for the case of triple stars (see e.g. \citealt{Shappee2013}; \citealt{Toonen2016}).  Similar effects have been studied in the case of planets in binaries using $N$-body simulations \citep{Kratter2012}.  These works, however, have focused on the impact that wind-driven mass loss may have on the dynamic of these systems. 

Before achieving hydrostatic equilibrium (i.e. zero main sequence or ZAMS for short), stars experience a relatively slow contraction. Depending on the stellar mass, the radius of a pre-main-sequence star may vary by a factor of up to 3 while ``traveling'' along the Hayashi track \citep{Harpaz1994}.  The time of these changes are of the order of the thermal timescale, which for the case of a $0.6\, M_{\odot}$ star is about $220$ Myr. Other changes in the stellar structure, relevant to their interaction with potential companions, also occur during these phases.  Thus for instance, the internal mass distribution and its response to tidal effects, which affect the interchange of angular momentum, also change while the star is contracting (see e.g. \citealt{Zuluaga2016}). On the dynamical side and depending on the relative orbital inclination of the binary and the planet (and to a less extent, on the initial values of inner eccentricty and periastron argument, see e.g. \citealt{Bataille2018}), high-amplitude excitations of the planetary eccentricity will happen \citep{Valtonen2006}. During these eccentricity excursions, the planet may reach temporarily small periapse distances and strong tidal interaction with the host star are triggered. Since tidal torques are very sensitive to stellar and planetary radius \citep{Murray1999}, the outcome of the dynamical evolution may be significantly different than in the case when we assume that those radius are constant or vary in time.  

We hypothesize that including changes in stellar and planetary radius into the dynamics will significantly impact the outcome of the early dynamical evolution of planets in binaries, especially in S-type systems.  Since the orbital ``architecture'' of these systems, their survival and stability, depend on what happen at the beginning of their life, including stellar and planetary evolution will have an impact on our interpretation of the observations.  This paper is also aimed to (partially) test this hypothesis. For the sake of conciseness, we will focus here on testing the effect of changes in stellar radius and left the effects of the evolution of stellar interior structure and planetary radius, for a future paper.

As we will show, we can identify a limited region in the space of inner semi-major axis and eccentricity, where the inclusion of changes in radius alters significantly the timescales of evolution.

This paper is structured as follows. In Section \ref{sec:marco_teorico} we present the conservative, non-conservative and relativistic formalisms that constitute the building blocks of the complete secular theory.  Section \ref{sec:stellar_astrophysics} presents some relevant aspects of the pre-main-sequence stellar evolution with an special focus on the timescales involved. In Section \ref{sec:numerical_experiments} we show and discuss the results of four numerical experiments we performed to test our hypothesis. Section \ref{sec:discussion} is intended to discuss the impact that other effects could have in the secular dynamics of these systems, the limitations and prospect of our models.

\section{Secular theory of hierarchical three-body systems}
\label{sec:marco_teorico}

The secular theory of hierarchical three-body systems have been widely developed and presented in literature (see eg. \citealt{Harrington1968,Ford2000,Naoz2013,Carvalho2016,Correia2016} and references there in).  For the sake of completeness, we reproduce in this section the most relevant results of the theory, paying special attention on presenting explicit expressions of the relevant formulas. Many of these formulas are absent in literature, due (partially) to the complexity of the mathematical involved expressions. In an effort to make all our results fully reproducible, we also refer the reader to Section \ref{sec:appendix_1} for further details of the mathematical expressions.



\section{Secular theory of hierarchical three-body systems}
\label{sec:marco_teorico}

The secular theory of hierarchical three-body systems have been widely developed and presented in literature (see e.g. \citealt{Harrington1968,Ford2000,Naoz2013,Carvalho2016,Correia2016} and references there in).  For the sake of completeness, we reproduce in this section the most relevant results of the theory, paying special attention on presenting explicit expressions of the relevant formulas. Many of these formulas are absent in literature, due (partially) to the complexity of the mathematical involved expressions. In an effort to make all our results fully reproducible, we also refer the reader to \autoref{sec:appendix_1} for further details of the mathematical expressions.



\subsection{Conservative motion}
\label{sec:conservative_motion}

In the absence of dissipative forces and relativistic effects, the Hamiltonian for the orbital motion of a general three body system is \citep{Harrington1968}\footnote{We have adopted the convention of defining the Hamiltonian as the negative of the mechanical energy to agree with the notation of previous works (e.g. \citealt{Ford2000}; \citealt{Naoz2013}; \citealt{Carvalho2016}).}:

\begin{equation}
\label{eq:01}
\begin{split}
\displaystyle 
\Hcal \equiv & \Hcal_{\mathrm{kep},1} + \Hcal_{\mathrm{kep},2} + \Hcal_{\mathrm{int}}\\
= & \frac{\mathcal{G}\m \mm}{2\ain} + 
\frac{\mathcal{G}(\m+\mm)\mmm}{2\aout} +\\
& + \frac{\mathcal{G}}{\aout}\sum_{n=2}^{\infty} \alpha^n M_n\bigg(\frac{\rin}{\ain}\bigg)^n\bigg(\frac{\aout}{\rout}\bigg)^{n+1}
P_n(\cos \Phi)
\end{split}
\end{equation}

\begin{figure*}
\centering
\includegraphics[width=0.8\textwidth]{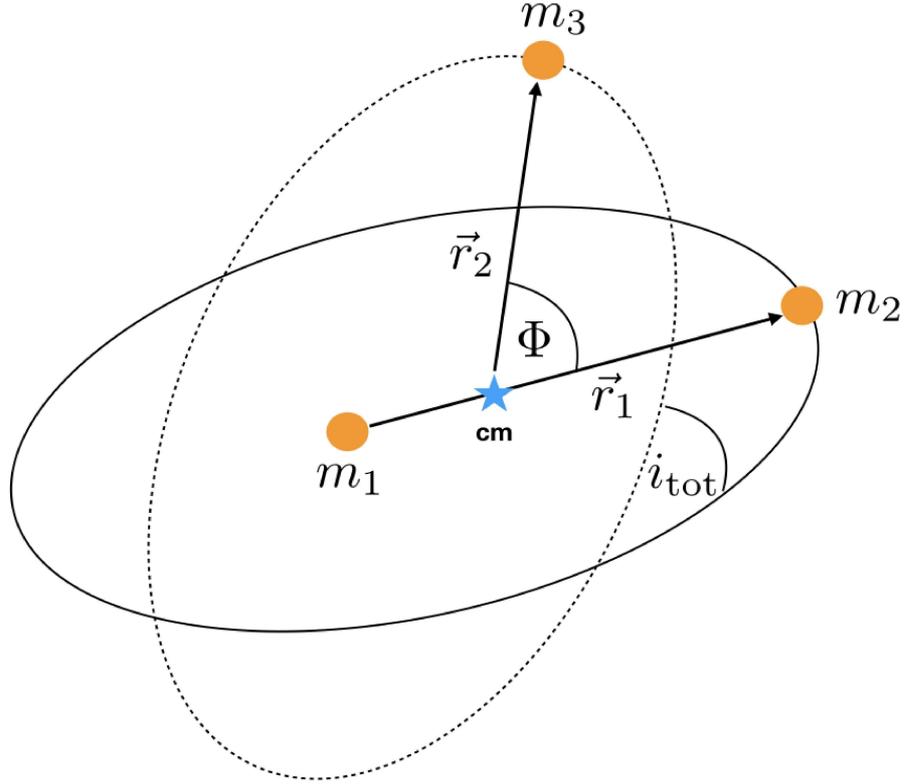}
\caption{The geometry (not to scale) of a hierarchical three body system showing the relative position vectors $\vec{r}_1, \vec{r}_1$ and their relative orientations. Since in general, $\itot \neq 0$, the angle $\Phi$ does not lie on any of the orbital planes. The blue star marks the position of the center of mass of the inner system.}
\label{fig:figura_1}
\end{figure*}

Here $M_n = \m \mm \mmm [\m^{n-1}-(-\mm)^{n-1}]/(\m+\mm)^n$, $P_n$ is the Legendre polynomial of degree $n$ and $\Phi$ is the angle between the two Jacobian coordinates $\vec{r}_1\equiv \vrin$, $\vec{r}_2\equiv \vrout$ (see Figure \ref{fig:figura_1}) and $\alpha \equiv \ain/\aout$ is the ratio of the semi major axis of the inner (star 1 and planet) and the outer system (star 1-planet and star 2). Hereafter, quantities with subindex $1$ and $2$ will be referred to the inner and outer orbits respectively.  To avoid confusion with other dynamical quantities, the gravitational constant will be written as $\mathcal{G}$. 

\vspace{0.2cm}

The first two terms of the Hamiltonian ($\Hcal_{\mathrm{kep},1}$, $\Hcal_{\mathrm{kep},2}$) correspond to the contribution of the relative Keplerian motions of $\mm$ around $\m$, and that of $\mmm$ around the center of mass of $\m$ and $\mm$, respectively. The third term, $\Hcal_{\mathrm{int}}$ correspond to the interaction Hamiltonian, which accounts for the perturbative effect of the outer orbit on the inner one.  The strength of the latter term is proportional to the relative size $\alpha$ of the inner and outer orbits.  If $\alpha$ is small enough, namely $\alpha<1/2$, secular perturbation techniques can be used \citep{Correia2016}. In this hierarchical regime, high order terms in the summation can be neglected. If the only term retained is that corresponding to $n=2$, the approximation is made to {\em quadrupole level}. When summation is expanded up to $n=3$, we have the {\em octupole} approximation. 

Since the Hamiltonian does not depends explicitly on time, we can use Hamilton-Jacobi theory to find a better set of canonical variables \citep{Goldstein}.  We will denote the new generalized coordinates as $l_j,h_j,g_j$ (where $j=1,2$ are indexes for the inner and outer system) and they are a function only of the initial conditions and time.  The corresponding conjugated momenta will be $L_j,H_j,G_j$, and they are all constants of motion. 

A particularly useful set of canonical coordinates for this system, fulfilling the previous conditions, are the Dalaunay elements \citep{Valtonen2006}:

\begin{alignat}{2}
\hspace{0.4in} l_j&=M_j,& \qquad L_j&=m_{r,j}\sqrt{\mu_ja_j}\label{eq:lM}\\
h_j&=\Omega_j,& \qquad H_j&=m_{r,j}\sqrt{\mu_ja_j (1-e_j^2)}\cos I_j\label{eq:hO}\\
g_j&=\omega_j,& \qquad G_j&=m_{r,j}\sqrt{\mu_ja_j (1-e_j^2)}\label{eq:gw}
\end{alignat}

\noindent where $\mu_j={\cal G}{\cal M}_j$, ${\cal M}_j$ is the total mass of each of the systems (${\cal M}_1=m_1+m_2$, ${\cal M}_2={\cal M}_1+m_3$), $m_{r,j}$ are the reduced masses: $m_{r,1}=m_1m_2/\mathcal{M}_1$, $m_{r,2}=m_3\mathcal{M}_1/\mathcal{M}_2$ and $M_j,\Omega_j,\omega_j$ are the mean anomalies, the longitudes of the ascending nodes and the arguments of the periapse with respect to an inertial whose fundamental plane is the Laplace plane of the three body system. 

In terms of these elements, the octupolar Hamiltonian is now written: 

\begin{equation}
\label{eq:02}
\begin{split}
\displaystyle \Hcal =& \frac{\beta_1}{2 L_1^2} + \frac{\beta_2}{2 L_2^2}
+ 8\beta_3 \frac{L_1^4}{L_2^6}\bigg(\frac{\rin}{\ain}\bigg)^2\bigg(\frac{\aout}{\rout}\bigg)^3 (3\cos^2 \Phi -1) \\
& + 2\beta_4 \frac{L_1^6}{L_2^8}\bigg(\frac{\rin}{\ain}\bigg)^3\bigg(\frac{\aout}{\rout}\bigg)^4 (5\cos^3 \Phi -3\cos \Phi)
\end{split}
\end{equation}

\noindent with $\beta_i$ being auxiliar functions of the particle masses, whose explicit expressions are presented in Appendix \ref{sec:appendix_1}.

All the varying quantities in the Hamiltonian, namely $\Phi$, $\rin$ and $\rout$, implicitly depend the generalized-mean anomalies $l_j$ (see equation \ref{eq:lM}) which are related to the true anomalies $f_j$.

In the ``natural'' (non-inertial) reference frame of each orbit ($x_j$ axis pointing to the periapse and $z_j$ axis being perpendicular to the orbital plane), the relative position of each system is simply \citep{Murray1999}:
$\mathbf{r}_{j,\mathrm{orb}}=r_{j,\mathrm{orb}}\mathbf{\hat{r}}_{j,\mathrm{orb}}$, with $\mathbf{\hat{r}}_{j,\mathrm{orb}}=(\cos f_j,\sin f_j,0)$ and:

\begin{equation}
\label{eq:6}
\displaystyle r_{j,\mathrm{orb}} = \frac{a_j(1-e_j^2)}{1+e_j\cos f_j}
\end{equation}


The position vector of each body in the inertial system will be:

\begin{equation}
\mathbf{r}_{j,\mathrm{in}}=\mathbf{M}(\Omega_j,I_j,\omega_j)\mathbf{r}_{j,\mathrm{orb}}
\end{equation}

\noindent with $\mathbf{M}(\Omega_j,I_j,w_j)=\mathbf{M}(h_j,I_j,g_j)$ an Eulerian $3\times 3$ rotation matrix ($I_j$ is the inclination of each orbit relative to the Laplace plane). Here, subscripts ``orb" and ``in" emphasizes that the quantity is measured in the orbital or inertial reference system, respectively. In the following derivation $\mathbf{M}^\intercal$ will be the tranpose matrix. 

Using this conventions, the relative angle $\Phi$ will be:

\begin{equation}
\begin{split}
\cos \Phi &= \mathbf{\hat{r}}_{2,\mathrm{in}} \cdot \mathbf{\hat{r}}_{1,\mathrm{in}} \\ &= \mathbf{\hat{r}}_{1,\mathrm{orb}}^\intercal \mathbf{M}^\intercal(h_1,I_1,g_1)  \mathbf{M}(h_2,I_2,g_2) \mathbf{\hat{r}}_{2,\mathrm{orb}}
\end{split}
\end{equation}

\noindent or explicitly, in terms of the Delaunay elements:

\begin{equation}
\label{eq:8}
\begin{split}
\displaystyle \cos \Phi =& \sin(f_2+g_2)\big[A_1\sin(f_1+g_1) + A_3\cos(f_1+g_1)\big] - \\
&\cos(f_2+g_2)\big[A_2\sin(f_1+g_1) - A_4\cos(f_1+g_1)\big]
\end{split}
\end{equation}

\noindent where we have introduced auxiliary functions $A_i(h_j,I_j,g_j)$ ($i=1,2,3,4$), whose explicit expressions are in Appendix \ref{sec:appendix_1}. 

When equations (\ref{eq:6}) and (\ref{eq:8}) are replaced in (\ref{eq:02}), the octupolar Hamiltonian becomes an explicit function of the true anomalies of both orbits. 

To construct a secular theory, we need to get rid from the Hamiltonian those rapidly oscillating variables. This is done with aid of the Von Zeipel transformation \citep{Marchal2012}, namely, if the system does not exhibit mean motion resonances, the dynamics could be described with the doubly-averaged interaction Hamiltonian:

\begin{equation}
\displaystyle \langle\langle \Hcal \rangle \rangle = \frac{1}{4 \pi^2} \int_{0}^{2\pi}\int_{0}^{2\pi} dl_1 dl_2 \Hcal_{\mathrm{int}}
\end{equation}

The resulting doubly-averaged Hamiltonian expanded to octupolar order reads\footnote{It is interesting to notice that the integration on the inner orbit is easier to be done if we instead of the true anomaly $l_1$, use the eccentric anomaly as the integration variable.}: 

\begin{equation}
\label{eq:full-hamiltonian}
\begin{split}
\displaystyle \langle\langle \Hcal \rangle \rangle \approx & \frac{2\beta_3}{\big(1-e_2^2\big)^{3/2}}\frac{L_1^4}{L_2^6}\bigg[-2(2 + 3e_1^2) +3\big(1-e_1^2\big)\big(K_1^2+K_4^2\big)+\\
& +3\big(1+4e_1^2\big)\big(K_2^2+K_3^2\big)\bigg] + \frac{15 \beta_4}{16 \big(1-e_2^2\big)^{5/2}}\frac{L_1^6}{L_2^8}\times \\
&\times\Bigg\{e_1e_2\bigg[4K_2 \big(4+3e_1^2\big)-5\big(3+4e_1^2\big)K_2\big(K_2^2+K_3^2\big)-\\
&5\big(1-e_1^2\big)\big(2K_1 K_3 K_4+3K_1^2K_2 + K_4^2 K_2\big)\bigg]\Bigg\}
\end{split}
\end{equation}

\noindent where explicit expressions for the auxiliary functions $K_i(h_j,I_j,g_j)$ ($i=1,2,3,4$) are also provided in Appendix \ref{sec:appendix_1}.  This is the first time that an analytical formula for this Hamiltonian without any simplification is provided explicitly in literature. 

This expression can be greatly simplified if we have in mind that when working in the Laplace plane $\Delta h = h_1-h_2 = 180^\circ$ \citep{Valtonen2006} a condition called ``elimination of the nodes''.  We need to be very cautious with this step. As discussed in \cite{Naoz2013}, this procedure is not correct because by applying it, we are reducing the total volume of the phase space over which are we will construct the variational trajectories to derive the equations of motion through Hamilton's principle. The elimination of the nodes directly in the Hamiltonian leads to erroneous conclusions like the independent conservation of the $z-$component of the angular momentum of both orbits: $H_j=G_j\cos I_j$ (\citealt{Kozai1962}). To derive the equations of motion for inner and outer inclinations, we rely in the geometric properties arising from the conservation of the \textit{total} angular momentum as explained in Appendix \ref{sec:appendix_1}. Fortunately, for the orbital elements whose defining equation of motion does not depends on the partial differential operator with respect to $h_i$, we can use the simplified Hamiltonian. Therefore, once this caution has been made, we can express the ``node-eliminated" form of the secular Hamiltonian as: 

\begin{equation}
\label{eq:final_hamiltonian}
\begin{split}
\displaystyle &\langle\langle \Hcal \rangle \rangle = \frac{\beta_3}{2\big(1-e_2^2\big)^{3/2}}\frac{L_1^4}{L_2^6}\Bigg\{\big(2+3e_1^2\big)\big(1+3\cos 2\itot\big)+ \\
& +30e_1^2\cos 2g_1 \sin^2\itot\Bigg\} + \frac{15 \beta_4 e_1e_2}{64 \big(1-e_2^2\big)^{5/2}}\frac{L_1^6}{L_2^8}\times \\
&\times\Bigg\{70e_1^2 \cos 2g_1 \sin^2 \itot \big(\cos g1 \cos g2 + \cos \itot \sin g_1 \sin g_2\big) + \\
& + \bigg(6-13e_1^2+5\big(2+5e_1^2\big)\cos 2\itot \bigg) \cos g_1 \cos g_2 + \\
& + \bigg(5\big(6+e_1^2\big)\cos 2\itot - 7\big(2-e_1^2\big)\bigg)\cos \itot \sin g_1 \sin g_2 \Bigg\} 
\end{split}
\end{equation}

Once the secular Hamiltonian has been derived, the secular equations of motion are given by the canonical Hamilton's equations: 

\begin{alignat}{2}
\hspace{0.2in} 
\frac{dL_j}{dt}&=\frac{\partial \secHcal }{\partial l_j},& \qquad \frac{dl_j}{dt}&=-\frac{\partial \secHcal }{\partial L_j}\\
\frac{dG_j}{dt}&=\frac{\partial \secHcal }{\partial g_j},& \qquad
\label{eq:relativistic}
\frac{dg_j}{dt}&=-\frac{\partial \secHcal }{\partial G_j}\\
\frac{dH_j}{dt}&=\frac{\partial \secHcal }{\partial h_j},& \qquad \frac{dh_j}{dt}&=-\frac{\partial \secHcal }{\partial H_j}
\end{alignat}

All this procedure produces a set of 8 (non-trivial) first order, nonlinear and coupled differential equations. The explicit expression of the equation of motion can be found in Appendix \ref{sec:conservative_motion} where we have followed the same notation of \cite{Naoz2013}. 

\subsection{Non-conservative motion}
\label{sec:tides}

As orbital evolution develops, the Lidov-Kozai effect generates high-amplitude oscillations of the inner eccentricity and mutual inclination \citep{Lidov1962,Kozai1962}. Depending on initial conditions, the eccentricity values reached can be high enough to produce close encounters at periastron passages, which in turn may trigger strong tidal interactions between the inner bodies. 

Tides have a direct impact on the orbital elements of the inner system and on the rotational properties of $m_1$ and $m_2$. On the other hand, since we are interested only on those systems which are ``hierarchically'' stable, i.e. systems where the hierarchy of the mean distances are satisfied along the evolution, we can safely neglect the tidal effects experimented by the outer body. 

Recently, many works have addressed the dynamics of tidal interaction with the aid of sophisticated rheological models
(e.g \citealt{Efroimsky2009}; \citealt{Ferraz-Mello2013}; \citealt{Correia2014}). Others have applied those models to the context of exoplanets dynamics (e.g \citealt{Makarov2012}; \citealt{Cuartas2016}). In this work, and for the sake of simplicity, we implemented the classical tidal formalism of \citet{Hut1981} and \citet{Eggleton1998}. 

In the kind of S-type systems we are studying here, the planet ($m_2$) raises a tidal bulge on its host star ($m_1$). Moreover, since the star is rotating (very fast at early ages), it also has a rotational distortion (that for simplicity we will express only to quadrupolar order).  The total two-body acceleration will be then given by:

\begin{equation}
\label{eq:twobeq}
\displaystyle \ddot{\vec{r}} = -\frac{\mathcal{G}(m_1+m_2)}{r_1^3}\vec{r}_1 + \vec{a}_{\mathrm{TF}} + \vec{a}_{\mathrm{QD}} 
\end{equation}

\noindent where the induced non-gravitational tidal $\vec{a}_{\mathrm{TF}}$ and quadrupolar-deformation $\vec{a}_{\mathrm{QD}}$ accelerations are given by \citep{Eggleton1998}:

\begin{eqnarray}
\label{eq:quad_accel}
\vec{a}_{\mathrm{TF}} & = & -\frac{C}{t_{Vs} r_1^{10}}\Big(3\vec{r}_1 \vec{r}_1\cdot \dot{\vec{r}}_1 + (\vec{J}_1-\vec{\Omega}_s r_1^2)\times \vec{r}_1\Big)\\
\vec{a}_{\mathrm{QD}} & = & \frac{m_2A_s}{m_r}\bigg(\frac{5(\vec{\Omega}_s\cdot \vec{r}_1)^2\vec{r}_1}{2r_1^7} + \nonumber\\
& & - \frac{\Omega_s^2 \vec{r}_1}{2r_1^5} - \frac{(\vec{\Omega}_s\cdot \vec{r}_1) \vec{\Omega}_s}{r_1^5} -  \frac{6\mathcal{G}m_2 \vec{r}_1}{r_1^8}\bigg)
\nonumber
\end{eqnarray}

\noindent where $m_r$ is the reduced mass of the inner system and $\vec{\Omega}_s$ is the rotational velocity vector of the star ($m_1$) whose orientation is completely arbitrary in this formalism. 

Hereafter, quantities labeled with subscripts $s$ and $p$ correspond to properties of the star and the planet, respectively.  

The tidal component of the acceleration in Equation (\ref{eq:twobeq}) depends on the angular momentum of the inner system:

\begin{equation*}
\displaystyle \vec{J}_1 = \sqrt{\mathcal{G}(m_1+m_2)a_1(1-e_1^2)} \hat{z}_1 ,
\end{equation*}

\noindent on the viscous timescale $t_{\mathrm{V1}}$ (which is assumed constant) and a rheological constant $C$, which is defined as:

\begin{equation}
\displaystyle C =\frac{9}{2} \bigg(\frac{1+2k_s}{2k_s m_1 R_s}\bigg)^2 m_2(m_1+m_2) 
\end{equation}

\noindent where $k_s$ is the apsidal motion constant (a measure of the quadrupolar deformability). This constant is related to the Love number of the star by $k_{L,s} = 2k_s$ \citep{Naoz2016}. 

From Equation (\ref{eq:quad_accel}), it is clear that tidal acceleration has a strong dependence  on the instantaneous separation of the inner bodies $r_1$, $a_{\rm TF}\propto r_1^{-10}$, which make its effect very relevant at short distances.

On the other hand the quadrupolar deformation contribution\footnote{It is worth noticing that the quadrupolar dissipative force arises in the regime when the mechanical energy is dissipated at a rate proportional to the square of the rate of change of the quadrupolar moment tensor\citep{Eggleton1998}.} is proportional to the factor $A_s=2k_sR_s^5$ which also depends strongly on the stellar radius.

After those forces are included in the equations of motion for $m_1$ and $m_2$, the next step is to derive the averaged (secularized) effect of them on the dynamical evolution of the inner system. For further details of the secularization processes we refer the reader to the work of \cite{Eggleton1998}. 

Here we just summarized the contribution of these dissipative forces to the inner secular dynamics. The complete set of secular equations that rules the orbital and rotational evolution due to the dissipative effects are given by explicit expressions in Appendix \ref{sec:appendix_1}. 

The magnitude of the tidal interaction on the orbital dynamics of the inner system, is better illustrated using the so-called tidal friction timescale for the star (\citealt{Eggleton1998}; \citealt{Naoz2016}):

\begin{equation}
\label{eq:tfA}
\displaystyle t_{F,s}=\frac{t_{V,s}}{9}\bigg(\frac{a_1}{R_s}\bigg)^8\frac{m_1^2}{(m_1+m_2)m_2}\frac{1}{(1+2k_s)^2}
\end{equation}

This quantity is a measure of the efficiency of energy dissipation due to tides. In general, the smallest $t_{Fs}$, the more efficient is the dissipation and therefore tidal evolution takes place more rapidly. 

Please notice the strong dependence of this quantity on stellar radius. This dependence implies, for instance, that an increment in the star's size by a factor of two, increases the rate of tidal dissipation by a factor of $256$.  This sensitivity on the stellar radius is precisely what motivate us to study the effect of stellar evolution in the dynamics of these systems.   

    \subsection{General relativity correction} 
    \label{sec:GR}
    
    Due to the excursions of the inner eccentricity to high values, relativistic effects are expected to have a major  impact, especially on the evolution of the argument of periastron of the inner orbit (periastron precession). 
    
    In order to include this effect, we implemented the following  Post-Newtonian correction to the Hamiltonian \citep{Richardson1988}:
    
    \begin{equation}
    \label{eq:H_gr}
    \displaystyle \mathcal{H}_{\mathrm{GR}} = \gamma_1 \vec{P}_1^4 + \gamma_2 \frac{\vec{P}_1^2}{r_1} + \gamma_3 \frac{(\vec{r}_1\cdot \vec{P}_1)^2}{r_1^3}+ \gamma_4 \frac{1}{r_1^2} 
    \end{equation}
    
    \noindent with $\vec{v}_1$ being the relative velocity of the inner system and $\vec{P}_1$ the specific (relativistic) momentum, defined as: 
    
    \begin{equation}
    \displaystyle \vec{P}_1=\vec{v}_1+\frac{1}{c^2}\left(4\gamma_1 (\vec{v}_1\cdot \vec{v}_1)\vec{v}_1 + \frac{2\gamma_2}{r_1} \vec{v}_1 + \frac{2\gamma_4}{r_1^3}(\vec{r}_1\cdot \vec{v}_1)\vec{r}_1 \right) 
    \end{equation}
    
    \noindent with $c$ the speed of light. In Appendix \ref{sec:appendix_1} we provide explicit expressions for the auxiliary functions $\gamma_i$. 
    
    If we approximate the momentum to first order in $v_1/c$, i.e. $\vec{P}_1 \approx \vec{v}_1$, the relativistic doubly-averaged Hamiltonian correction will have error terms of the order $\mathcal{O}(v_1^4/c^4)$ and can be written as: 
    
    \begin{equation}
    \begin{split}
    \label{eq:H_gr_2}
    \displaystyle \langle\langle \mathcal{H}_{\mathrm{GR}} \rangle \rangle =& \gamma_1 \langle\langle\vec{v}_1^4 \rangle \rangle + \gamma_2 \BDlangle\frac{\vec{v}_1^2}{r_1} \BDrangle + \gamma_3 \BDlangle \frac{(\vec{r}_1\cdot \vec{v}_1)^2}{r_1^3} \BDrangle \\
    &+ \gamma_4 \BDlangle \frac{1}{r_1^2} \BDrangle 
    \end{split}
    \end{equation}
    
    For a comprehensive explanation of the secularization process of this Hamiltonian, we refer the reader to the work of \cite{Migaszewski2009}. the final secularized relativistic Hamiltonian will be: 
    
    \begin{equation}
    \displaystyle \langle \langle \mathcal{H}_{\mathrm{GR}} \rangle \rangle = -\frac{3\mu^4 \beta^5}{c^2 L_1^3 G_1} 
    \end{equation}
    
    
    \noindent and, from Equation (\ref{eq:relativistic}), the corresponding general relativity contribution to equation of motion for the argument of periastron will be:
    
    \begin{equation*}
    \displaystyle \bigg(\frac{dg_1}{dt}\bigg)_{\mathrm{GR}} = -\frac{\partial \langle \langle \mathcal{H}_{GR} \rangle \rangle}{\partial G_1} = \frac{3\mu^4\beta^5}{c^2 L_1^3 G_1^2}
    \end{equation*}
    
    \noindent that can be expressed explicitly in terms of the eccentricity and semi-major axes, as: 
    
    \begin{equation}
    \displaystyle \bigg(\frac{dg_1}{dt}\bigg)_{\mathrm{GR}} = \frac{3\mu^{3/2}}{c^2 a_1^{5/2}(1-e_1^2)}
    \end{equation}
    
    \bigskip
    
    In summary, the dynamical evolution of a hierarchical triple system obeys equations of motion with contributions coming from a ``conservative'' Hamiltonian, a non conservative contribution only associated to the interaction of $m_1$ and $m_2$, and last but not least, a relativistic correction on the argument of the pericenter of the inner orbit $g_1$. 
    
    Among all these effects, the non-conservative one strongly depends on the stellar radius.  This dependency, and as we will show later, can not be neglected when studying the evolution of the system during those phases when the radius of the star changes in timescales similar to that of the orbital dynamics.  We will focus here on the early phases (pre-main-sequence) when the radius decreases by a factor of 2-3 in low-mass stars.

\section{Pre-main-sequence stellar evolution} 
\label{sec:stellar_astrophysics}

Stars are formed in overdensed regions of a primogenitor molecular cloud. The progressive release of gravitational potential energy leads to the formation of a protostar which is still embedded inside the cloud making it visible only as an infrared source. When the cloud is dispersed and the protostar has reached its final mass, a pre-main-sequence star was born \citep{Carroll2007}. 

For masses lower than $3\, M_{\odot}$, the radius and observed properties of the star (effective temperature, luminosity, surface gravity, etc.) evolves following what is called a Hayashi track \citep{Hayashi1966}.  The particular track followed by a star will depends on its mass and composition. 

\begin{figure}
\centering
\includegraphics[width=\columnwidth]{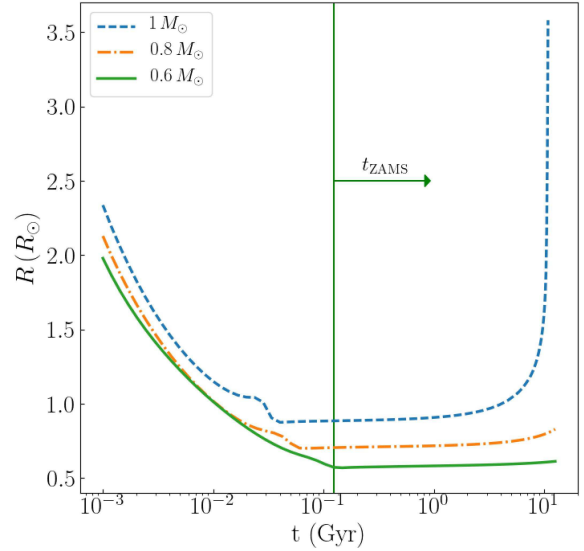}
\caption{Evolution of the stellar radius for three stars with masses equal to $0.6\, M_{\odot}$ (continuous green), $0.8\, M_{\odot}$ (dashed-dotted orange) and $1\, M_{\odot}$ (dashed blue). The vertical line marks the arrival to the zero age main sequence (ZAMS) of the $0.6\, M_{\odot}$ star.}
\label{fig:tres_radios}
\end{figure}

Along its Hayashi track the luminosity of the pre-main-sequence stars drops while maintaining a roughly constant effective temperature.  These changes are a by product of the sustained decrease in radius of the star (see Figure \ref{fig:tres_radios}). 

While in its Hayashi track, the star is characterized by having a fully convective interior with a thin radiative photosphere. However, due to the continuous increase in central temperature, the opacity of the core decreases leading to the formation of a radiative nucleus.  At this point the central region rapidly heats reaching temperatures high enough to trigger nuclear fusion of hydrogen-1 into helium-4 \citep{Harpaz1994}.  Hydrogen fusion provides enough thermal energy to guarantee hydrostatic equilibrium and the star reaches his final radius.  At this point the star is said to be a zero age main sequence star (ZAMS).  

The duration of the pre-main-sequence phase (from the arrival to the Hayashi track to ZAMS) is of the order of the thermal (or Kelvin-Helmholtz) timescale \citep{Toonen2016}:

\begin{equation}
\label{eq:thermal_time}
\displaystyle \tau_{\mathrm{th}} = \frac{\mathcal{G}m_s^2}{R_s L_s}
\end{equation}

\noindent where here $L_S$ is the stellar instantaneous luminosity. Assuming the ZAMS values of $R_s$ and $L_s$ of a $0.6\, M_{\odot}$ mass star, $\tau_{\mathrm{th}} \approx {\cal O}(100$\;{\rm Myr}) which is consistent with the results of precise simulations.  Since $R_s\propto L_s^{1/2}$ (for almost constant effective temperature) and $L_s\propto m_s^{3.5}$ (\citealt{Carroll2007}; \citealt{Demircan1991}), the timescale of pre-main-sequence collapse is $\tau_{\rm th}\propto m_s^{-3.25}$.  This implies that more massive stars will have a considerably shorter pre-main-sequence, but also that low mass stars will spend several hundred of Myrs in this evolutionary phase.


In Figure \ref{fig:tres_radios} we see the evolution of stellar radius for three solar-metallicity stars with masses $1,\, 0.8$ and $0.6$ solar masses. The results were obtained from the \texttt{PARSEC v1.2}\footnote{\url{https://people.sissa.it/~sbressan/}} evolutionary tracks \citep{Bressan2012}. 

\vspace{0cm}

Vertical green line represents the onset of the zero age main sequence (ZAMS) phase in the $0.6\, M_{\odot}$ case. At the left of the vertical line, we can see how the radius evolves in the pre-main-sequence phase, changing by a factor of about $3$ during a time interval similar to that predicted by equation (\ref{eq:thermal_time}).      

All what has been said here correspond to the evolution of an isolated star. In a binary system, the pre-main sequence phase could be altered by the presence of a companion, specially in the case of compact S-type systems \citep{Toonen2016}. We further discuss the potential implications of these effects in Section \ref{sec:discussion}.

\section{Numerical experiments}
\label{sec:numerical_experiments}

In order to test the effect that the change in stellar radius has in the dynamical evolution of S-type binaries, we have devised 4 numerical experiments:



\begin{itemize}

\item \textbf{Experiment 1:} Secular dynamical evolution of a real binary planet, namely HD 80606 b. This system has been already studied in literature and we can use it for verification purposes. 

\item \textbf{Experiment 2:} A generic S-type system assuming constant stellar radius.  This experiment will allow us to illustrate the typical behavior of some dynamical quantities and define the ``proxies'' we will use to evaluate the effect of stellar radius evolution. 

\item \textbf{Experiment 3:} A systematic exploration of the parameter space (initial orbital parameters). We are interested on to identify the initial values $a_1$, $e_1$ and $\itot$ for which the role of a variable stellar radius is more important.

\item \textbf{Experiment 4:} A new case of study. With the initial conditions identified in Experiment 3 we fully simulate a hypothetical system.
\end{itemize}

To have a configurable set of numerical scripts to perform these experiments, but also to  allow other researchers to reproduce our results, we developed and published an open C package, \texttt{SecDev3B}\footnote{\url{https://github.com/bayronportilla/SecDev3B}}.  Provided a set of initial conditions, \texttt{SecDev3B} solves the complete set of differential equations that rules the secular evolution of a hierarchical system of three bodies. A detailed description of this package is included in the {\tt README} of the {\tt GitHub} repository. 

\subsection{Experiment 1: revisiting the case of HD 80606b}
\label{sec:experiment_1}

Here, we revisited the long-term dynamical evolution of one of the most studied binary planets, HD 80606b.  For this purpose, we solve the secular equation of motion coming from the conservative (expanded to octupolar degree) and the relativistic terms of the Hamiltonian as well as the non-conservative equations of motion.  For this particular experiment and in order to compare our results to previous ones, we assume constant stellar radius. 

HD 80606b is an eccentric ($e = 0.93$) hot Jupiter orbiting the G5-type star HD 80606 which is located at $190$ light-years away in the Ursa Major constellation. The host star is member of a binary system with the HD 80607.  The average separation between the components of the binary system is $\sim 1200$ AU \citep{Moutou2009}. 

Radial velocity measurements and the analysis of the Rossiter-McLaughlin anomaly has suggested that the planet is in a prograde, highly inclined orbit around its host star. 

\citet{Wu2003} has suggested that the observed configuration is the result of the perturbative effect of the outer star. They performed simulations of the system using a secular theory expanded to quadrupolar order including the same non-conservative effects described in Section \ref{sec:tides} and a general relativity correction to the inner argument of pericenter. 

With the set of initial conditions in Table 1 of \citealt{Wu2003}, a solution was obtained.  Their results suggest that the system underwent Lidov-Kozai oscillations during the first $2.7$ Gyr.  Using the observed values of eccentricity and semi-major axis, they also conclude that the planet is not experiencing this effect anymore. Independently other authors \citep{Fabrycky2007,Correia2011} studied the system with analogues results.

We use our package \texttt{SecDev3B} to reproduce these results.  In Table \ref{tab:table_1} we present the initial conditions we use in our simulations.  We prepare two different simulations in this case: 1) using the expansion of the conservative Hamiltonian up to quadrupolar order; and 2) using the expansion up to the octupolar term.  The latter numerical experiment was never attempted before. 

In order to see why the octupolar terms may become relevant in this particular case, we should notice that the octupolar term in equation (\ref{eq:final_hamiltonian}) is proportional to the factor $\beta_4\sim (\m - \mm)$ (see Appendix \ref{sec:appendix_1}).  This term can only be neglected when both members of the inner binary have similar masses.  In this case, however, since $\m/\mm \sim 150$, the octupolar term may become relevant.  Our numerical results confirm this expectation.

\begin{table*}
	\centering
	\caption{Initial conditions of each experiment described in this work. Quantities marked with (-) takes a series of values depending on the grid properties described in Section \ref{sec:experiment_3}. }
	\label{tab:table_1}
	\begin{tabular}{lllll} 
		\hline
		Property& Experiment 1 & Experiment 2 & Experiment 3 & Experiment 4\\
		\hline
		Type & S-type & S-type & S-type & S-type\\
		Protoype & HD 80606b & Fictitious & Fictitious & Fictitious\\
		$a_1; \, a_2$ (AU) & 5;\,1000 & $2.5;\, 300 $ & -;\, 50 & 1;\, 50 \\
        $e_1; \, e_2$ & 0.1;\, 0.5 & $0.6;\, 0.3$ & -;\, 0.3 & 0.35;\, 0.3\\
        $\itot$ (deg) & 85.6 & $80$ & - & 80  \\
        $g_1; \, g_2$ (deg) & 45;\, 0 & $0;\, 0$ & 0;\, 0 & 0;\, 0\\
        $\m; \, \mm; \, \mmm$ ($M_{\odot}$) & 1.1;\, 0.00744;\, 1.1 & $1;\, 0.001;\, 0.667$ & 0.6;\, 0.01;\, 0.6 & 0.6;\, 0.01;\, 0.6\\
        $\R; \, \RR$ ($R_{\odot}$) & $1;\, 0.1$  & $1;\, 0.1$ & 0.59;\, 0.1 & 0.59;\, 0.1\\
        $\ProtA; \, \ProtB$ (day) & $20;\, 0.416$  & $20;\, 1$ & 20;\, 1 & 20;\, 1\\
        $\tvA; \, \tvB$ (years) & $50;\, 0.001$  & $50;\, 0.001$ & 50;\, 0.001 & 50;\, 0.001\\
        $k_s; \, k_p$ & $0.014;\, 0.255$  & $0.014;\, 0.255$ & 0.014;\, 0.255 & 0.014;\, 0.255\\
        $\kappaA; \, \kappaB$ & $0.08;\, 0.25$  & $0.08;\, 0.25$ & 0.08;\, 0.25 & 0.08;\, 0.25\\
		\hline
	\end{tabular}
\end{table*}

\begin{figure*}
\centering
\includegraphics[width=0.9\textwidth]{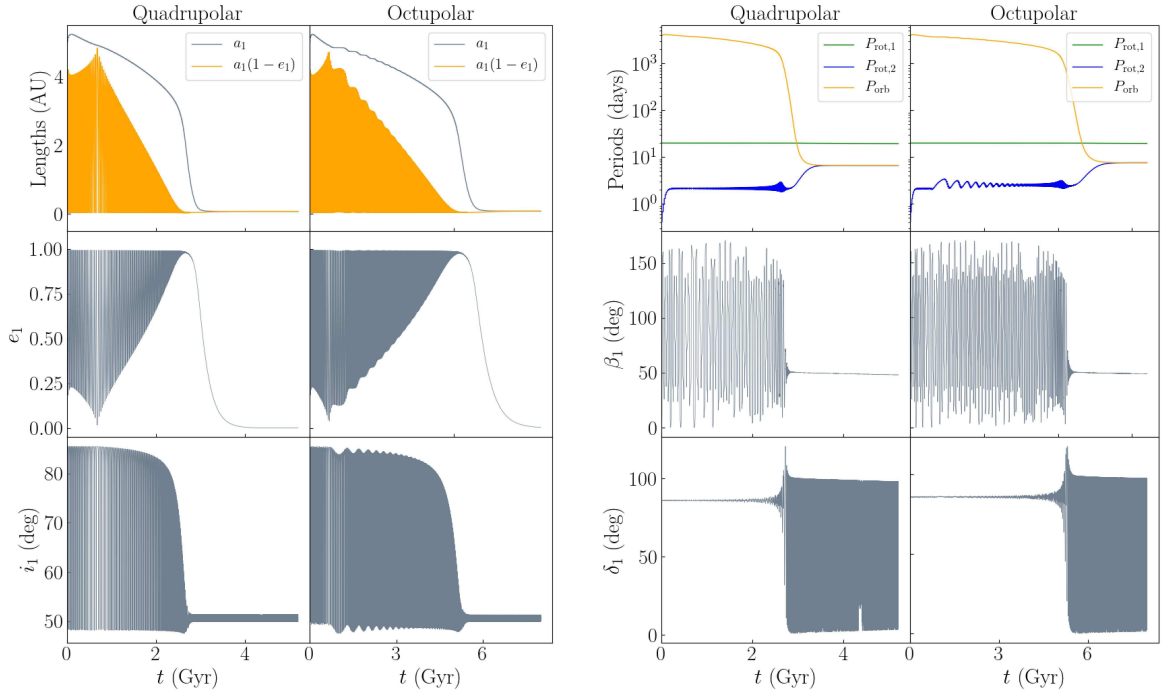}
\caption{The possible evolution of HD 80606b planet considering the initial conditions for experiment 1 given in table \ref{tab:table_1}. The simulation was performed both with the quadrupolar and octupolar approximations. Although the final values of each quantity are identical in both approximations, the timescales of evolution are completely different. In this plot, $\beta_1$ and $\delta_1$ represent the stellar obliquities with respect to the inner orbit and to the Laplace plane, respectively. Our results to quadrupolar order are identical to those reported by \citealt{Wu2003}, \citealt{Fabrycky2007} and \citealt{Correia2011}.}
\label{fig:HD80606b}
\end{figure*}

In Figure \ref{fig:HD80606b} we present the result of our experiment with HD 80606b. The results to quadrupolar order are identical to those reported in previous works \citep{Wu2003,Fabrycky2007,Correia2011}. This confirm that our package is well-suited for this problem.

\vspace{0.5cm}

As a by product of this test, we find that the inclusion of the octupolar term leads to a significant modification of the dynamical evolution timescales. In particular, the time required to suppress the Lidov-Kozai oscillations, or in other terms, the time required for the oscillations in eccentricity to reach an amplitude less than $1\%$ of its initial value (see second row in Fig. \ref{fig:HD80606b}) at quadrupolar-order is about $2.7$ Gyr whereas when the octupolar term is included this time becomes about $5.3$ Gyr.

This significant increase of the dynamical timescale is triggered by the emergence of longer ``plateaus'' in the inner semi-major axis and eccentricity oscillations (steps in the yellow shaded lines and in the semi-major axis curve in the first row of Fig. \ref{fig:HD80606b}). When the system is subject to strong Lidov-Kozai oscillations, the inner semi-major axis is temporarily ``captured'' into quasi-stable states (where the value of the semi-major axis becomes constant for tens of Myrs).  This ``stable'' periods are observed both when the quadrupolar and the octupolar terms are included.  However they are much longer in the latter case, delaying the orbit decay. It is also worth noticing that the reached final values for all dynamical quantities are the same under both approximations.           

Octupolar terms in the Hamiltonian implies the inclusion of extra harmonics into a finite series development. In the octupolar integration, we can notice the appearance of additional oscillation frequencies for each quantity plotted.  This additional frequencies shift the solution and again produces larger timescales. 

The actual impact this ``octupolar delay'' may have in our understanding of planets in binaries is diverse. Thus for instance, it could affect our expectation of the time spent for a planet in the habitable zone or the time required to reach or leave the unstable region (\citealt{Kane2013}; \citealt{Mason2015}).     

Should the octupolar order be included in every simulation?  Not necessarily.  In the case of S-type binary planets, since the inner system will be always composed by two bodies of masses considerably different, the octupolar term should always be included (unless the outer eccentricity is zero). In the case of P-type systems, where the components of the inner system have similar masses, this condition is relaxed reducing the contribution of the octupolar term to the total energy (see Equation \ref{eq:final_hamiltonian}). 

\subsection{Experiment 2: a generic S-type system}
\label{sec:experiment_2}

\begin{figure}
\centering
\includegraphics[width=\columnwidth]{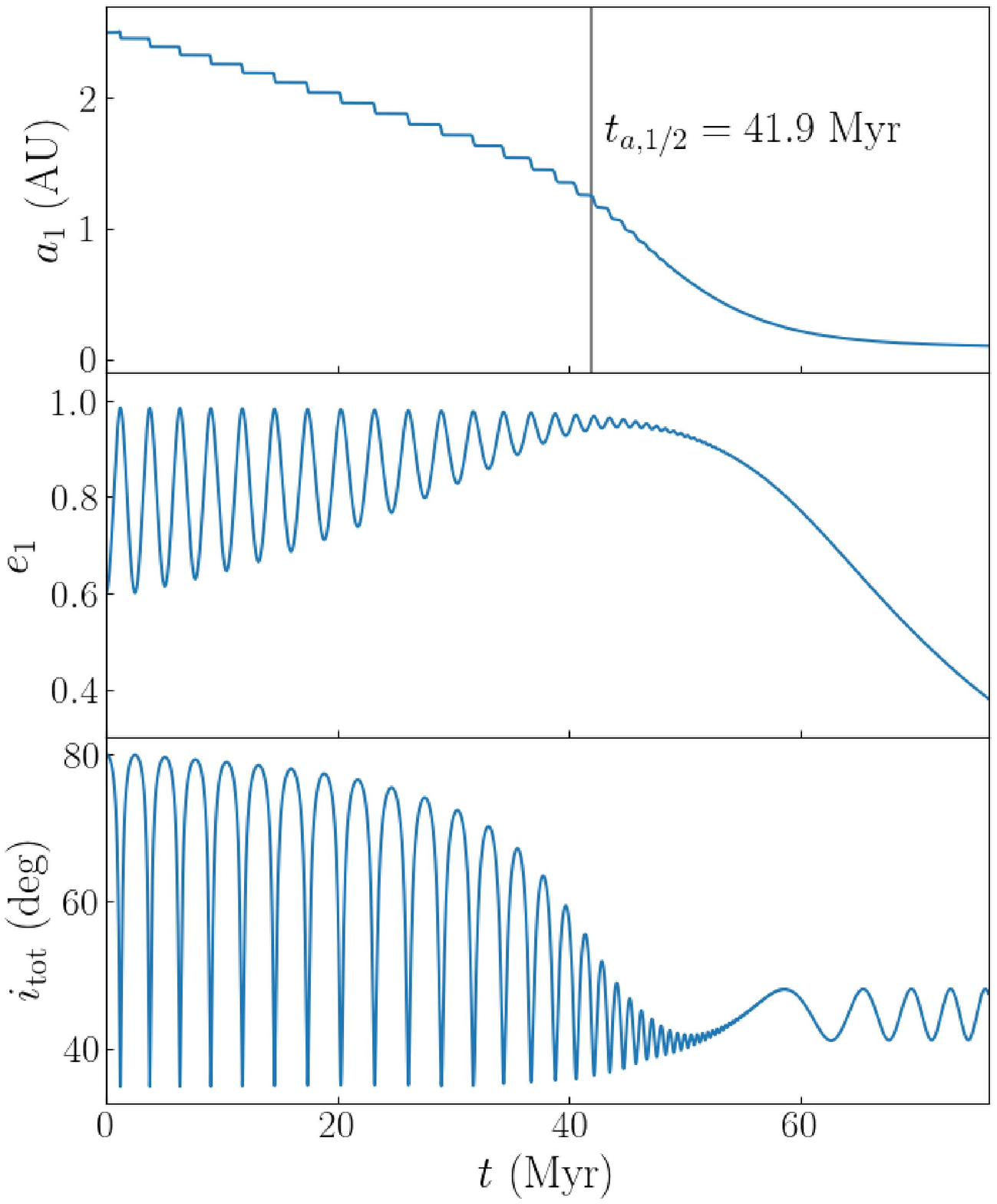}
\caption{Secular evolution of a hypothetical S-type system including non-conservative and general relativity effects. The vertical line in the upper row marks the time $\tahalf$ when the semi-major axis is half of the initial one.}
\label{fig:ejemplo}
\end{figure}

Once we verify that our code runs properly, we want to identify the characteristic features of the secular evolution of our hierarchical systems.  For this purpose we simulate a hypothetical S-type binary starting with the initial conditions given in Table \ref{tab:table_1}. In Figure \ref{fig:ejemplo} we present the result of our simulation in this case.

\vspace{0.5cm}

The first noticeable feature is the ``ladder-shaped'' evolution of the semi-major axis.  This behavior was already noticed in Experiment 1, however here it is easier to identify the possible reasons of this behavior.  The almost sudden drop in the semi-major axis are correlated with excursions to high-eccentricity values in the inner system. Thus, for instance, the first maximum of $e_1$ is $0.985$ reached after $1.2$ Myr, which is the same time when the first drop in $a_1$ happens. We must keep in mind that in this secular formalism the only phenomena able to modify the value of the inner semi-major axis are the non-conservative effects.  
The correlation between the drops in $a_1$ and the excursion of $e_1$ becomes evident if we have in mind that the magnitude of the tidal acceleration goes as $r^{-10}$ (see equation \ref{eq:quad_accel}), and therefore, it is much bigger in each periapse passage.  When eccentricity gets its highest value, the periapse distance gets very small (for a given value of the semi-major axis). For instance, at $t=1.2$ Myr, the periastron distance is just $\sim 0.04$ AU. This distance is small enough to trigger a strong tidal interaction which is the responsible of the sudden drop in the value of $a_1$.


Our results also evidence the strong anticorrelation between the inner eccentricity and the mutual inclination.  This anticorrelation arise from the interchange of angular momentum between both the inner and outer orbits. This is precisely what we call the Lidov-Kozai mechanism (\citealt{Lidov1962}; \citealt{Kozai1962}). 

In our experiments, we have identified that the time required for the semi-major axis to reduce to the half of its initial value ($\tahalf$):

\begin{equation}
\displaystyle a_1(t=\tahalf) = \frac{1}{2}a_1(t=0)
\end{equation}

\noindent is a good proxy of the inner orbit evolution.


We could also define similar time scales for other quantities such as eccentricity or inclination. However, since the evolution in eccentricity is mainly controlled by the conservative terms of the Hamiltonian, the introduction of the evolution in the stellar radius will not modify significantly the value of those other proxies of dynamical evolution.


\subsection{Experiment 3: the effect of stellar evolution}
\label{sec:experiment_3}

In order to test how the evolution of the stellar radius impact the timescales of orbital evolution in S-type binaries, we should explore the space of all possible initial conditions.  If we consider only the orbital evolution the space to explore has 11 dimensions (eight possible initial conditions and three masses). The size of the parameter space is significantly increased if we include the rotational parameters (moment of inertial, radius, rotational rates, etc.)  

Since this paper is intended only to test the hypothesis that the evolution of stellar radius may have a significant effect on the dynamical evolution of hierarchical systems, we restrict our selves to ``subspaces'' of the whole parameter space.  In particular we will restrict to two sections  $a_1\, \mathrm{vs.}\, e_1$ and $a_1\, \mathrm{vs.}\, i_{\mathrm{tot}}$. For this purpose we fixed all the remaining relevant physical and orbital parameters.  

For the $a_1$ vs. $e_1$ subspace we set initial mutual inclination in $80^\circ$ (which is as large as that observed in HD 180606b). For the $a_1$ vs. $\itot$ subspace the initial value of the inner eccentricity was set in $0.6$.

\subsubsection{The $a_1$ vs. $e_1$ space}
\label{sec:experiment_3_1}

The exploration of this subspace begins with the construction of a $10 \times 10$ grid of $(a_1,e_1)$ values. The semi-major axes cover a range between $0.85$ AU to $4.0$ AU with a constant separation of $0.35$ AU. On the other hand, inner eccentricity goes from $0.18$ to $0.9$ in steps of $0.08$. 

Each point in the grid represents an individual simulation with initial conditions given by the coordinates of the point. The remaining initial parameters are summarized in Table \ref{tab:table_1}. 

A first set of simulations was made maintaining a constant value of the stellar radius. The decay time of the semi-major axes ($\tahalf$) was recorded for each simulation. Next, a second set of $100$ simulations was ran with the same initial conditions, but this time letting the radius to change according to the predicted evolution of a $0.6\, M_{\odot}$ star (see Fig. \ref{fig:tres_radios}). Once again, the decay time for $a_1$ was also recorded. 

\begin{figure*}
\centering
\includegraphics[width=0.8\textwidth]{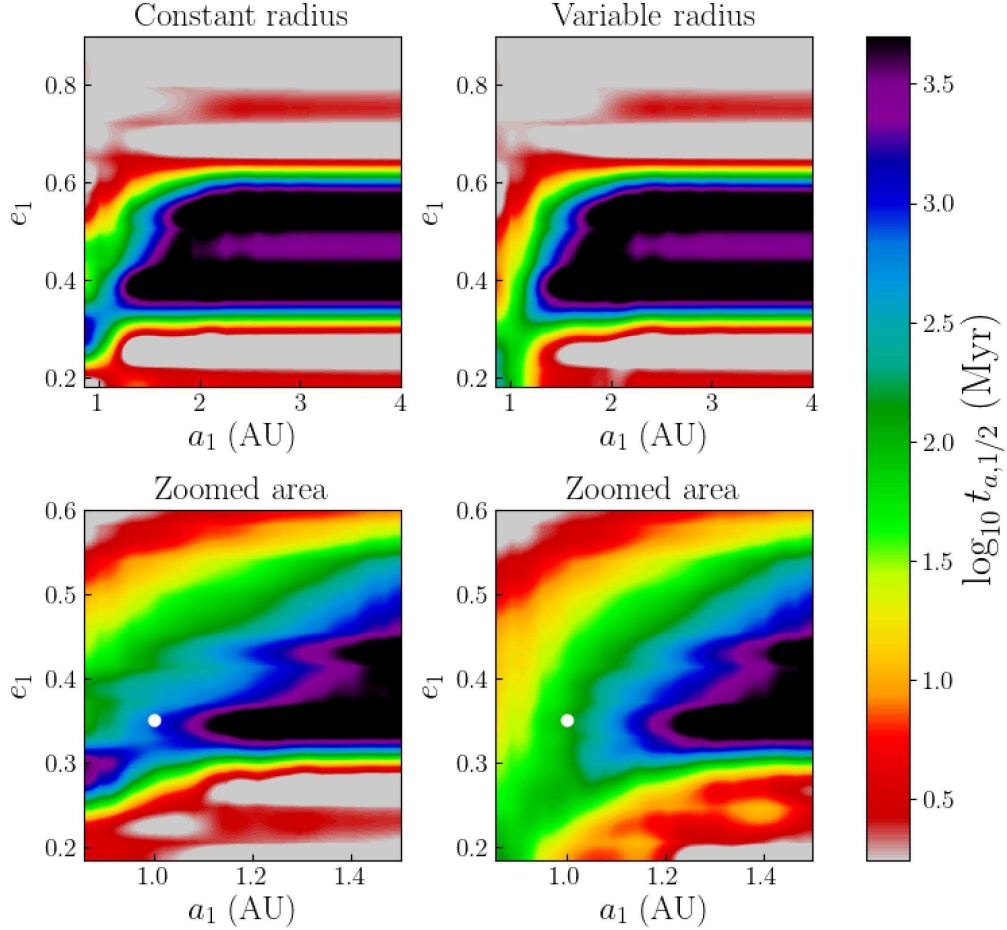}
\caption{Contour plots of the decay times for the inner semi-major axis in the $a_1$ vs. $e_1$ space, in the case of constant stellar radius (left column) and variable stellar radius (right column). Upper row show a coarse grid covering a larger range of the parameters.  The lower row shows a zoom into an identified area of interest.}
\label{fig:ae_space-a_half}
\end{figure*}

If after $5000$ Myr, the inner semi-major does not match the decay condition, namely  $a_1(t=5000)\leq a_1(t=0)/2$, the simulation was stopped and the value assigned to $\tahalf$ was $5000$ Myr. Once all the simulations were performed, each point in the grid gets a value of $\tahalf$. In other words, we obtained a set of discrete sample points of the function $\tahalf = \tahalf(a_1,e_1)$. Finally, we interpolate the discrete values and obtain contour plots of $\tahalf(a_1,e_1)$.  The result of our simulations are shown in the contours of Figure \ref{fig:ae_space-a_half}.

\vspace{0cm}

There are noticeable differences in decay times for low initial semi-major axes. This is not surprising since the effect of variable radius is relevant for the non-conservative interaction, which has a strong dependence on the average inner system separation (Equation \ref{eq:tfA}). 

To analyze more closely this region, we made a ``zoom'' in the area defined by $0.8 \text{ AU } \leq a_1 \leq 1.5 \text{ AU }$ and $0.18 \leq e_1 \leq 0.6$. The results are shown in the panels of the lower row of Figure \ref{fig:ae_space-a_half}. Differences in decay times are again quite evident. Thus, for instance, for the selected initial conditions $(a_1,e_1)=(1,0.35)$ (white dots), the decay time with constant stellar radius is of the order of $700$ Myr. Including the effect of a variable stellar radius, the decay time drops to $150$ Myr. 

To explain this behavior, we should notice that at the beginning of the orbital evolution in the case of variable stellar radius, the initial value of this quantity was $\sim 2 \Rsun$.  This is almost 3 times bigger than the ZAMS radius which was precisely the value used in the constant stellar radius case.  Naturally, the higher the stellar radius, higher is the energy dissipation due to tides (see Equation \ref{eq:tfA}). Details of the evolution of a hypothetical system having this initial conditions are presented and discussed in Section \ref{sec:experiment_4}.  
  
\subsubsection{The $a_1$ vs. $i_{\mathrm{tot}}$ space}
\label{sec:experiment_3_2}

\begin{figure*}
\centering
\includegraphics[width=0.8\textwidth]{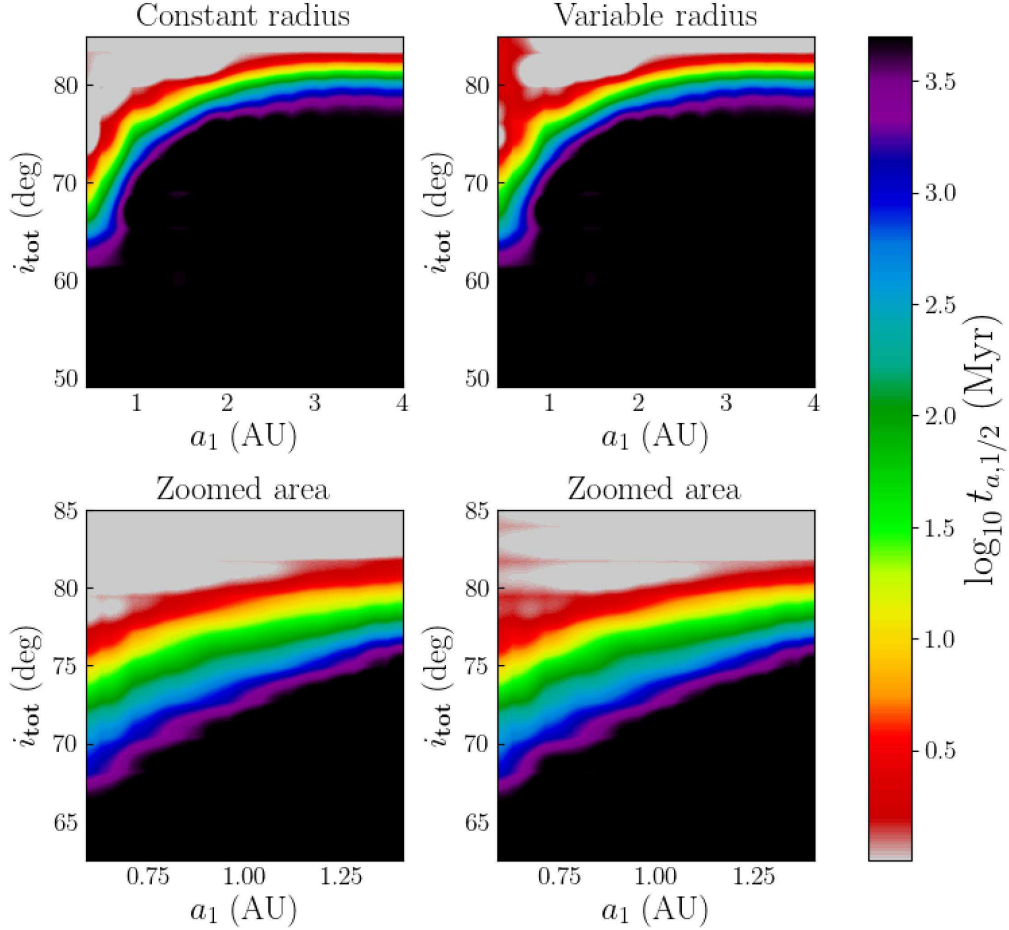}
\caption{Decay times for the inner semi-major axis in the $a_1$ vs. $\itot$ space. Left panels show the times obtained with integrations at constant radius. Right panels show the same but in the variable radius regime. Decay times appears to be insensitive to bulk variations in this subset of the parameter space.}
\label{fig:ai_space-a_half}
\end{figure*}

We repeat the previous procedure but now in the $a_1$ vs. $i_{\mathrm{tot}}$ subspace. The grid has the same range and periodicity in $a_1$ than before and the inclination was varied from $50^\circ$ to $85^\circ$ with a constant step of $4^\circ$.  A zoomed area was selected, spanning $0.6 \text{ AU } \leq a_1 \leq 1.4 \text{ AU}$ and $62.5^\circ \leq \itot \leq 85^\circ$.  The initial value of the inner eccentricity was fixed in $0.6$ while the rest of binary parameters are in Table \ref{tab:table_1}.  The resulting contour plots of $\tahalf(a_1,i_{\mathrm{tot}})$ are shown in Figure \ref{fig:ai_space-a_half}. 

\vspace{0cm}

Our proxy is now less sensitive to the selected set of parameters.  At low inclinations, the inner semi-major axis stays stable for a long-time (it does not decay).
Lidov-Kozai oscillations have very small-amplitude and are unable to bring the planet close enough to the star for the tidal interaction to have a real impact on the timescale of orbital evolution.

More importantly, at this specific eccentricity and range of inclinations, the orbital decay times are much less sensitive to the evolution in stellar radius.

\subsection{Experiment 4: a hypothetical evolving S-type system}
\label{sec:experiment_4}

\begin{figure*}
\centering
\includegraphics[width=0.6\textwidth]{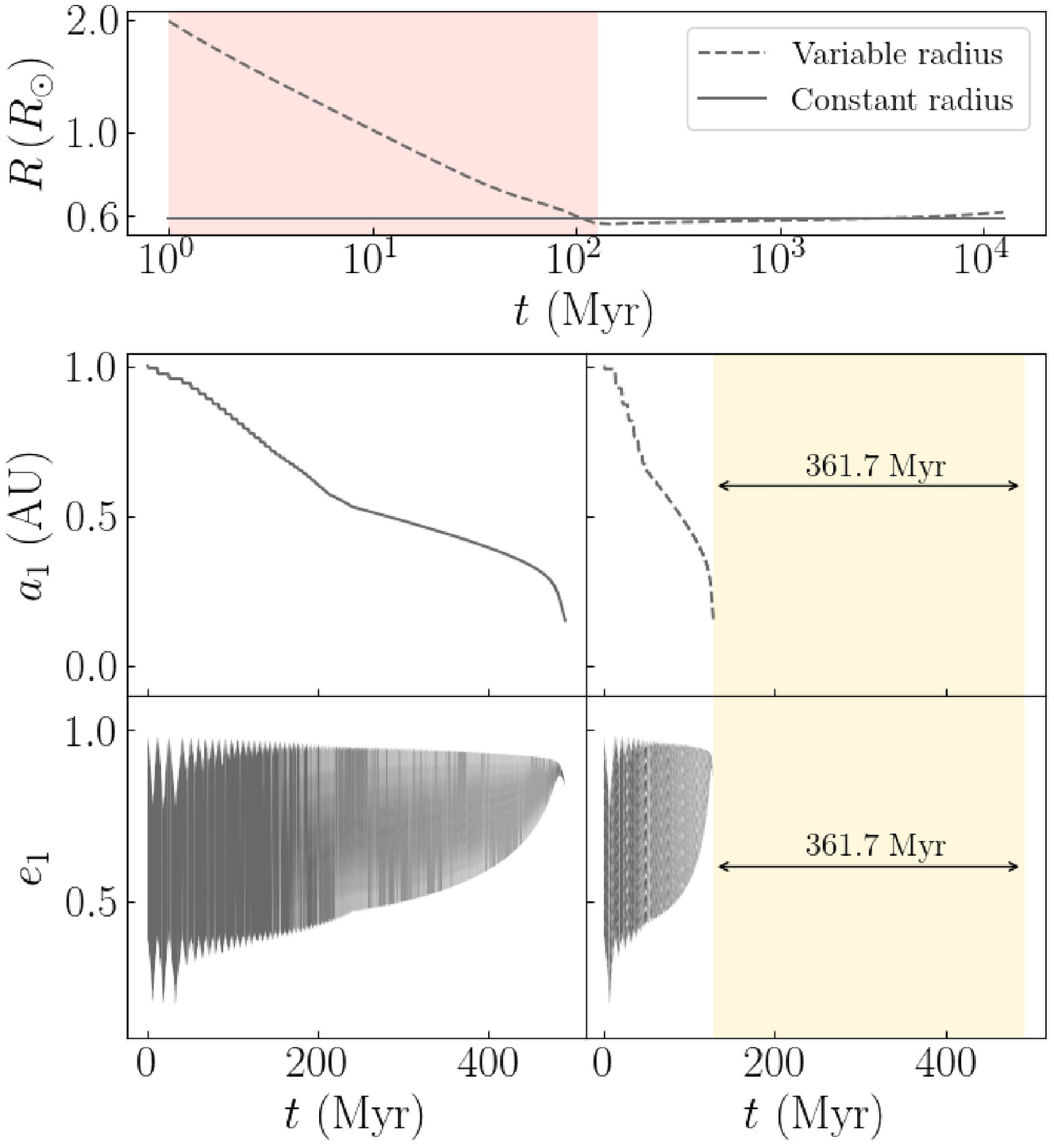}
\caption{Secular evolution of a hypothetical S-type system with a $0.6 \, M_{\odot}$ inner star.  Solid lines correspond to the orbital evolution in the case of constant stellar radius.  Dashed lines are for a star whose radius is contracting during the pre-main sequence phase (shaded area in the upper panel.  The difference between the duration of the high-amplitude Lidov-Kozai oscillations in both cases is highlighted in the second column of the middle and lower panels.}
\vspace{0.5cm}
\label{fig:huge_impact}
\end{figure*}

For our last experiment, we simulate the orbital evolution of a system having the properties of a selected spot (white point in Figure \ref{fig:ae_space-a_half}) in the $a_1$ vs. $e_1$ subspace.  Initial conditions for this system are enumerated in Table \ref{tab:table_1}.  Results of this experiment are summarized in Figure \ref{fig:huge_impact}.

Our planet belongs to a ``twin'' binary of 0.6 $\Msun$ stars which orbit each other in an eccentric orbit ($e_2=0.3$) with a semi-major axis of 50 AU.  The planet is initially on a mildly eccentric orbit ($e_1=0.35$) at $1$ AU from its host star. 

According to stellar evolution models the pre-main-sequence of the host star takes around 200 Myr (see upper panel in Fig. \ref{fig:huge_impact}) during which the stellar radius varied from about $2\, \Rsun$ to $0.6\, \Rsun$ \citep{Bressan2012}. Interestingly, for the particular orbital properties, the timescales of stellar evolution are of the same order than the timescales of orbital evolution. 

It is significant the effect that an evolving stellar radius has in the timescale of dynamical evolution.  While in the case of a constant radius star (left panel in Fig. \ref{fig:huge_impact}) it takes almost 200 Myr for the orbit to decay at about half of its initial value, when the stellar evolution is turned on the time-scale is reduced to less than 100 Myr. Even more dramatic is reduction in 362 Myr in the case of an evolving star, of the total time required for the end of the high-amplitude Lidov-Kozai oscillations.

In both cases however, the orbital properties at the end of the simulation was almost identical, namely, $a_{1,\mathrm{end}} = 0.15$ AU, while the final eccentricity were similar but not identical ($e_{1,\mathrm{end}} = 0.83$ in the constant-radius case and  $e_{1,\mathrm{end}} = 0.86$ in the evolving one). 


\section{Discussion}
\label{sec:discussion} 

Stellar and planetary evolution in binaries does not happen in isolation.  In the presence of a binary companion, different physical processes may affect the formation and evolution not only of the planet but the stars itself. 

In the case of P-type circumbinary systems, the distance between both stars are in general relatively small. When the radius of one of the stellar member is larger than its Roche lobe, mass could be transferred to the companion. This phenomena is known as Roche lobe overflow (RLOF) and may occur during the protostellar evolution and even the pre-main-sequence phase \citep{Ivanova2015}.  When matter is transferred from one star to another, the hydrostatic and thermal equilibrium of the donor star are disturbed and the radius and timescales shown in Figure \ref{fig:tres_radios} are modified. 

To evaluate under which circumstances an early RLOF may affect pre-main-sequence evolution we need to estimate the Roche radius $R_{L1}$. For a binary system with a mass ratio $q=m_{s1}/m_{s2}$ it is given by \citep{Eggleton1983}:

\begin{equation}
\label{eq:roche_radius}
\displaystyle R_{L1} \approx 0.44 a_{\mathrm{bin}}\frac{q^{0.33}}{(1+q)^{0.2}}
\end{equation}

For the system we simulate in experiment 4 (see Section \ref{sec:experiment_4}), $q=1$ and $a_{\mathrm{bin}} = 50$ AU, the Roche radius is about $19$ AU. Therefore, even when the highest expected possible value of the stellar radius is assumed, it is less than $0.04\%$ the Roche radius. RLOF effects on stellar evolution in this systems can be safely neglected and we can assume that both stars evolve following the models of isolated stars.

Stars transfer angular momentum not only via tidal interactions with stellar and planetary companions.  Mass-loss and magnetic breaking are other mechanisms in action, and they could be very strong, especially during the early phases of stellar evolution \citep{Huang1963,Yakut2008,Zuluaga2016}. These effects may also impact the total angular momentum budget in the system affecting orbital evolution.

Mass-losses induced by strong stellar winds at the beginning of the life of a low mass-star, may have an impact on the evolution of the binary orbit itself.  Thus, for instance, For a $0.6\, M_{\odot}$ T-tauri star the mass loss rates are as large as $3.5\times 10^{-8} \, M_{\odot}/yr$ \citep{Ezer1971}.  If the ejection process is adiabatic and none of the wind-matter is accreted by the companion star, the average separation between both bodies changes according to \citep{Toonen2016}:

\begin{equation}
\label{eq:widen_no_acc}
\displaystyle \bigg(\frac{da_{\mathrm{bin}}}{dt}\bigg)_{\mathrm{no-acc}} = -\frac{\dot{m}_{s1}}{m_{s1}+m_{s2}}a_{\mathrm{bin}}
\end{equation}

\noindent when we use the notation adopted in Section \ref{sec:conservative_motion}, $a_{\rm bin}$ will be $a_1$ or $a_2$ depending on the nature of the system (S or P type). In this adiabatic case, the eccentricity of the binary orbit ($e_1$ or $e_2$ depending on the type) is not modified by the mass-loss.  


On the other hand, if the expelled material is additionally accreted by the companion star, there will be a transfer of angular momentum between them and binary eccentricity will not be constant any more (\citealt{Huang1963}; \citealt{Toonen2016}). 

None of these effects were taken into account in our simplified models.  This simple estimations however, contribute to reveal the complex role that stellar evolution may play at determining the orbital fate of planet in binaries.

Planetary formation is also an important factor in the dynamical evolution of planets in binaries.  It will not only determine the initial conditions at which our simulations start, but several key early processes may affect orbital evolution. Circumstellar (or circumbinary) disks may last between $0.1$ Myr to $10$ Myr \citep{Haisch2001,Pelupessy2013,Haghighipour2010}.  Although these times are relatively small with respect to the few hundreds of Myr involved in the orbital evolution of planet in binaries,  the presence of a (perturbed) disk during the first phases of orbital evolution should be taken into account.

\cite{Haghighipour2010} performed numerical experiments to study the effect on orbital evolution of an S-type planet when it was embedded in a circumstellar disk. They performed simulations of giant planets with and without accretion.  Their results showed that binary planets in this type of systems may migrate inwards and get their eccentricity excited both by the disk and the outer companion (see Fig. 6.5 of \citealt{Haghighipour2010}).  If the planet accretes gas from the disk the timescales of semi-major axis and eccentricity evolution, as well as their asymptotic values were significantly modified with respect to the non-accretion case.  




\section{Summary and conclusions}
\label{sec:conclusions}

We revisit in this paper the well-known problem of hierarchical three body systems applied to the case of planets in binaries.  For this purpose we deduce and provide the full general secularized Hamiltonian up octupolar order in Delaunay variables. In the equations of motion, we also included non-conservative (tidal interaction and quadrupolar deformation) and general relativity terms of the interaction between the inner bodies. To solve the equations under general initial conditions we developed an open source package that is properly documented and can be used to reproduce our results. During a given integration, \texttt{SecDev3B} is well suited to include or not the contributions from non-conservative and relativistics effects as desired by the user. It is also possible to explore how the change of the inner bodies' bulk properties affects the dynamics as those quantities can be treated as variables whose value will be interpolated from a provided data set. We apply this model to study the dynamical evolution of S-type binary planets.

Using this formalism we investigated the secular dynamics of the well-known binary planet HD 80606b.  We find that using a conservative Hamiltionian expanded to quadrupolar order our software reproduces properly the results found in literature.  However, and as a novel contribution, when we used the expansion to octupolar order, the evolutionary timescales were significantly modified.  This result highlight the importance to use high-order developments when dealing with hierarchical S-type binaries which does not match the requirement of having inner bodies of equal mass.

We then apply the model to study the role that the evolution of stellar radius during the pre-main-sequence phase could have in the secular dynamics of the system.  For this purpose we explore a small part of a $>$12-dimensional parameter space, in particular, subspaces of semi-major axes, eccentricities and mutual inclinations.  We found that in S-type systems around low-mass stars, with relative high inclinations ($\itot\ge 60\deg$), moderate eccentricities ($0.2\le e\le 0.4$) and for planets located around 1 AU (typical distances of the Habitable zones of the host stars), the changes experienced by the star during the first few hundreds of Myr alter significantly the dynamical evolution timescales.  In a particular case of study we found that a planet decay from an initial low eccentric orbit of 1 AU to a final eccentric orbit of 0.1 AU around its host star in a time almost 400 Myr shorter than that calculated assuming the star have a constant radius.

This first approach is still incomplete.  Other regions of the parameter space should be explored and non-trivial effects such as early stellar mass-loss, binary angular momentum interchange, planetary radius evolution, the role of a circumstellar disk during the first Myrs, among other should be considered in future developments of this exciting topic.

The more important conclusion derived from this work is that for assessing the fate and history of the orbital properties of planets in binaries, not only dynamical issues should be at play.  It is worth realizing that the evolution of the orbits of planets and stars in binaries, especially during the first hundred of Myrs, actually happens to evolving non-static objects.

\section*{Acknowledgements}

We are grateful with all the colleagues that through the discussion in seminars and presentations of this work in conferences directly and inderectly contribute to create its final form.  We especially thanks to Josué Cardoso dos Santos for his insightful comments.  B.P. was funded by the program {\em Jóvenes Investigadores e Innovadores 2016} of COLCIENCIAS through grant JIC 018-2017. He wants to thanks to the people in the {\em Vicerrectoría de Investigación} for his support during the program.








\onecolumn
\appendix

\section{Detailed mathematical expressions}
\label{sec:appendix_1}

In this appendix we will give explicit expressions for the auxiliary constants, functions and equation of motion mentioned through the paper. 

\subsection{Conservative components}
\label{sec:conservative_motion}

Quantities given in expressions \ref{eq:eqs_conservative_motion} appears explicitly in the expansion of the three body (non-secularized) Hamiltonian. In order to avoid confussion, please notice we have adopted the caligraphyc letter $\Gcal$ to represent the Cavendish gravitational constant.)

\begin{subequations}
\label{eq:eqs_conservative_motion}
\begin{align}
\displaystyle L_1 &= \frac{\m \mm}{\m + \mm}\sqrt{\Gcal(\m + \mm)a_1}\\
\displaystyle L_2 &= \frac{(\m \mm)\mmm}{\m + \mm + \mmm}\sqrt{\Gcal(\m + \mm + \mmm)a_2}\\
\displaystyle \beta_1 &= \Gcal\m \mm \frac{L_1^2}{a_1}\\
\displaystyle \beta_2 &= \Gcal(\m+\mm)\mmm \frac{L_2^2}{a_2}\\
\displaystyle \beta_3 &= \frac{\Gcal^2}{16}\frac{(\m+\mm)^7}{(\m+\mm+\mmm)^3}\frac{\mmm^7}{(\m \mm)^3}\\
\displaystyle \beta_4 &= \frac{\Gcal^2}{4}\frac{(\m+\mm)^9}{(\m+\mm+\mmm)^4}\frac{(\m-\mm)\mmm^9}{(\m \mm)^5} 
\end{align}
\end{subequations}

The literal expression of the doubly averaged Hamiltonian (before the node elimination) is cumbersome. To simplify it, we provide in the paper the complete expansion in terms of the auxiliary quantities  $K_i\, (i=1,2,3,4)$ that are a function of the orbital elements of the inner and outer orbits.  

\begin{subequations}
\begin{align}
\displaystyle A_1 &= \cos I_1 \cos I_2 \cos \Delta h + \sin I_1 \sin I_2 \\
\displaystyle A_2 &= \sin \Delta h \cos I_1 \\
\displaystyle A_3 &= \cos I_2 \sin \Delta h \\
\displaystyle A_4 &= \cos \Delta h \\
\displaystyle D_1 &= A_1 \sin g_2 - A_2\cos g_2 \\
\displaystyle D_2 &= A_3 \sin g_2 + A_4\cos g_2 \\
\displaystyle D_3 &= A_3 \cos g_2 - A_4\sin g_2 \\
\displaystyle D_4 &= A_1 \cos g_2 + A_2\sin g_2 \\
\displaystyle K_1 &= D_1 \cos g_1 - D_2\sin g_1 \\
\displaystyle K_2 &= D_1 \sin g_1 + D_2\cos g_1 \\
\displaystyle K_3 &= D_3 \cos g_1 + D_4\sin g_1 \\
\displaystyle K_4 &= D_4 \cos g_1 - D_3\sin g_1
\end{align}
\end{subequations}

The direct application of the Hamilton canonical equations to the  leads to the following 

The equation of motion derived from the doubly averaged Hamiltonian expanded to octupolar order are provided below.  Subindex ``orb" emphasizes the fact that the expression is valid only for the conservative (or orbital) regime.  To simplify the expression we have adopted the notation of \cite{Naoz2013}.

\begin{subequations}
\label{eq:orbital_evol}
\begin{align}
\begin{split}
\displaystyle \bigg(\frac{de_1}{dt}\bigg)_{\mathrm{orb}} =& C_2\frac{1-e_1^2}{G_1}(30e_1\sin^2 \itot \sin 2g_1) + C_3 e_2 \frac{1-e_1^2}{G_1}\big[30\cos \phi \sin^2 \itot e_1^2 \sin 2g_1 \\&- 10\cos \itot \sin^2 \itot \cos g_1 \sin g_2 (1-e_1^2) - \mathcal{A}(\sin g_1 \cos g_2 - \cos \itot \cos g_1 \sin g_2)\big]
\end{split}\\
\begin{split}
\displaystyle \bigg(\frac{de_2}{dt}\bigg)_{\mathrm{orb}} =& -C_3e_1\frac{1-e_2^2}{G_2}\big[10\cos \itot \sin^2 \itot (1-e_1^2)\sin g_1 \cos g_2 + \mathcal{A}(\cos g_1 \sin g_2 - \cos \itot \sin g_1 \cos g_2)\big] 
\end{split}\\
\begin{split}
\displaystyle \bigg(\frac{dg_1}{dt}\bigg)_{\mathrm{orb}} =& 
6C_2\bigg[\frac{1}{G_1}\bigg(4\cos^2 \itot + (5\cos 2g_1-1)(1-e_1^2-\cos^2 \itot)\bigg) + \frac{\cos \itot}{G_2}\big(2+e_1^2(3-5\cos 2g_1)\big)\bigg] \\ &-C_3e_2\bigg[e_1\bigg(\frac{1}{G_2} + \frac{\cos \itot}{G_1}\bigg)\big(\sin g_1 \sin g_2 \big(10(3\cos^2 \itot-1)(1-e_1^2)+A\big)-5\mathcal{B}\cos \itot \cos \phi\big) \\& -\frac{1-e_1^2}{e_1G_1}\big(\sin g_1 \sin g_2 10 \cos \itot \sin^2 \itot (1-3e_1^2)+\cos \phi (3\mathcal{A} - 10 \cos^2 \itot + 2)\big)\bigg]
\end{split}\\
\begin{split}
\displaystyle \bigg(\frac{dg_2}{dt}\bigg)_{\mathrm{orb}} =& 3C_2\bigg[\frac{2\cos \itot}{G_1}\big(2+e_1^2(3-5\cos 2g_1)\big)+\frac{1}{G_2}\bigg(4+6e_1^2+(5\cos^2 \itot-3)\big(2+e_1^2[3-5\cos 2g_1]\big)\bigg)\bigg] \\& +C_3e_1\bigg[\sin g_1 \sin g_2 \bigg(\frac{4e_2^2+1}{e_2G_2}10 \cos \itot \sin^2 \itot (1-e_1^2)- \\& e_2\bigg(\frac{1}{G_1}+\frac{\cos \itot}{G_2}\bigg)\big[\mathcal{A}+10(3\cos^2 \itot - 1)(1-e_1^2)\big]\bigg) + \cos \phi \bigg(5\mathcal{B} \cos \itot e_2 \bigg(\frac{1}{G_1} + \frac{\cos \itot}{G_2}\bigg) + \frac{4e_2^2 + 1}{e_2G_2}\mathcal{A}\bigg)\bigg]
\end{split}\\
\begin{split}
\displaystyle \bigg(\frac{dh_1}{dt}\bigg)_{\mathrm{orb}} =& 
-\frac{3C_2}{G_1\sin i_1}(2+3e_1^2-5e_1^2\cos 2g_1)\sin 2\itot \\& - C_3e_1e_2\bigg(5\mathcal{B}\cos \itot \cos \phi - \mathcal{A}\sin g_1\sin g_2 + 10(1-3\cos^2 \itot)(1-e_1^2)\sin g_1 \sin g_2\bigg)\frac{\sin \itot}{G_1 \sin i_1}
\end{split}\\
\begin{split}
\displaystyle \bigg(\frac{dh_2}{dt}\bigg)_{\mathrm{orb}} =& -\bigg(\frac{dh_1}{dt}\bigg)_{\mathrm{cons}}
\end{split}
\end{align}
\end{subequations}

As was explained in Section \ref{sec:conservative_motion}, the proper way to derive the equations of motion for the orbital inclinations is through the conservation law of the total angular momentum $\vec{G}_{\mathrm{tot}}$. For the sake of completeness, we must to strees the fact that this quantity is the vector summation of the orbital plus the spin angular momentum. However, the former component is by far bigger than the latter and we can work with the approximation: $\vec{G}_{\mathrm{tot}} \approx \vec{G}_1+\vec{G}_2$. The The magnitudes of the inner and outer orbital angular momentum are given by 

\begin{subequations}
\begin{align}
\displaystyle G_1 &= L_1\sqrt{1-e_1^2}\\
\displaystyle G_2 &= L_2\sqrt{1-e_2^2}
\end{align}
\end{subequations}

\noindent and the $z-$components are just:

\begin{subequations}
\begin{align}
\displaystyle H_1 &= G_1 \cos i_1\\
\displaystyle H_2 &= G_2 \cos i_2
\end{align}
\end{subequations}

The magnitude of the total angular momentum can be written in terms of those $z-$ components as  $G_{\mathrm{tot}} = H_1 + H_2$. The rate of change of those angular momenta are given by:

\begin{subequations}
\begin{align}
\begin{split}
\displaystyle \dot{G}_1 =& -C_2 30 e_1^2 \sin 2g_1 \sin^2 \itot + C_3 e_1 e_2 \big[-35 e_1^2 \sin^2 \itot \sin 2g_1 \cos \phi \\& + \mathcal{A}(\sin g_1 \cos g_2 - \cos \itot \cos g_1 \sin g_2)+10 \cos \itot \sin^2 \itot (1-e_1^2)\cos g_1 \sin g_2\big]
\end{split}\\
\begin{split}
\displaystyle \dot{G}_2 =& -C_3 e_1e_2 \big[\mathcal{A}(\cos g_1 \sin g_2 - \cos \itot \sin g_1 \cos g_2) + 10\cos \itot \sin^2 \itot (1-e_1^2)\sin g_1 \cos g_2\big]
\end{split}\\
\begin{split}
\displaystyle \dot{H}_1 =& \frac{G_1\dot{G}_1}{G_{\mathrm{tot}}} - \frac{G_2\dot{G}_2}{G_{\mathrm{tot}}} 
\end{split}\\
\begin{split}
\displaystyle \dot{H}_1 =& \frac{G_1\dot{G}_1}{G_{\mathrm{tot}}} - \frac{G_2\dot{G}_2}{G_{\mathrm{tot}}} 
\end{split}
\end{align}
\end{subequations}

In terms of those quantities, the equations of motion for the inclinations are: 

\begin{subequations}
\begin{align}
\displaystyle \bigg(\frac{di_1}{dt}\bigg)_{\mathrm{orb}} &= \frac{-\dot{H}_1+\dot{G}_1\cos i_1}{G_1\sin i_1} \\  
\displaystyle \bigg(\frac{di_2}{dt}\bigg)_{\mathrm{orb}} &= \frac{-\dot{H}_2+\dot{G}_2\cos i_2}{G_2\sin i_2}  
\end{align}
\end{subequations}

In previous equations, the following auxiliary quantities have been used:

\begin{subequations}
\begin{align}
\displaystyle &\mathcal{B} = 2+5e_1^2-7e_1^2\cos 2g_1\\
\displaystyle &\mathcal{A} = 4+3e_1^2-\frac{5}{2}B\sin^2 \itot\\
\displaystyle &\cos \phi = -\cos g_1 \cos g_2 - \sin g_1 \sin g_2 \cos \itot\\
\displaystyle &C_2 = \frac{\mathcal{G}^2}{16}\frac{(m_1+m_2)^7}{(m_1+m_2+m_3)^3}\frac{m_3^7}{(m_1 m_2)^3}\frac{L_1^4}{L_2^3 G_2^3}\\
\displaystyle &C_3 = -\frac{15}{16}\frac{\mathcal{G}^2}{4} \frac{(m_1+m_2)^9}{(m_1+m_2+m_3)^4} \frac{m_3^9(m_1-m_2)}{(m_1m_2)^5}\frac{L_1^6}{L_2^3G_2^5}
\end{align}
\end{subequations}

\subsection{Relativistic correction}

Many of the definitions given in this section are based in the notation of \cite{Richardson1988} and \cite{Migaszewski2009}. If we define $\sigma\equiv\m \mm/(\m+\mm)^2$ and $\mu\equiv G(\m+\mm)$. The auxiliary factors $\gamma_i$ appearing in equation (\ref{eq:H_gr_2}) are functions of the masses as follows:

\begin{subequations}
\begin{align}
\displaystyle \gamma_1 &= -\frac{1-3\sigma}{8c^2}\\
\displaystyle \gamma_2 &= -\frac{\mu(3+\sigma)}{2c^2}\\
\displaystyle \gamma_3 &= -\frac{\mu \sigma}{2c^2}\\
\displaystyle \gamma_4 &= \frac{\mu^2}{2c^2}
\end{align}
\end{subequations}

\subsection{Non-conservative motion}

We define the reduced mass of the inner system as $m_r = m_1m_2/(m_1+m_2)$ and the mean motion as $n=\sqrt{\mu/a_1^3}$. The equations of motion of the non-conservative part are written in terms of the following auxiliary functions of the inner eccentricity.   

\begin{subequations}
\begin{align}
\displaystyle f_1(e_1) &= \frac{1+15e_1^2/4+15e_1^4/8+5e_1^6/64}{(1-e_1^2)^{13/2}} \\
\displaystyle f_2(e_1) &= \frac{1+3e_1^2/2 + e_1^4/8}{(1-e_1^2)^5} \\
\displaystyle f_3(e_1) &= \frac{1+15e_1^2/2 + 45e_1^4/8 + 5e_1^6/16}{(1-e_1^2)^{13/2}} \\
\displaystyle f_4(e_1) &= \frac{1+3e_1^2 + 3e_1^4/8}{(1-e_1^2)^5}\\
\displaystyle f_5(e_1) &= \frac{1+9e_1^2/2 + 5e_1^4/8}{(1-e_1^2)^5}
\end{align}
\end{subequations}

Please notice that the definition of $f_i(e_1)$ differs from that adopted in \cite{Correia2011}.

The following auxiliary functions of the rheological parameters and rotational velocity of each body should be defined to write down the non-conservative components of the equation of motion.  

\begin{subequations}
\begin{align}
\displaystyle V_s &= \frac{9}{\tfA}\bigg(f_1(e_1)-\frac{11 \OmAz}{18n}f_2(e_1)\bigg)\\
\displaystyle V_p &= \frac{9}{\tfB}\bigg(f_1(e_1)-\frac{11 \OmBz}{18n}f_2(e_1)\bigg)\\
\displaystyle W_s &= \frac{1}{\tfA}\bigg(f_3(e_1)-\frac{\OmAz}{n}f_4(e_1)\bigg)\\
\displaystyle W_p &= \frac{1}{\tfB}\bigg(f_3(e_1)-\frac{\OmBz}{n}f_4(e_1)\bigg)\\
\displaystyle X_s &= -\frac{m_2 k_s R_s^5}{\mu n a_1^5}\frac{\OmAz \OmAx}{(1-e_1^2)^2} - \frac{\OmAy}{2n\tfA}f_5(e_1)\\
\displaystyle X_p &= -\frac{m_1 k_p R_p^5}{\mu n a_1^5}\frac{\OmBz \OmBx}{(1-e_1^2)^2} - \frac{\OmBy}{2n\tfB}f_5(e_1)\\
\displaystyle Y_s &= -\frac{m_2 k_s R_s^5}{\mu n a_1^5}\frac{\OmAz \OmAy}{(1-e_1^2)^2} + \frac{\OmAx}{2n\tfA}f_2(e_1)\\
\displaystyle Y_p &= -\frac{m_1 k_p R_p^5}{\mu n a_1^5}\frac{\OmBz \OmBy}{(1-e_1^2)^2} + \frac{\OmBx}{2n\tfB}f_2(e_1)\\
\displaystyle Z_s &= \frac{m_2 k_s R_s^5}{\mu n a_1^5}\Bigg(\frac{2\OmAz^2-\OmAy^2-\OmAx^2}{2(1-e_1^2)^2}+\frac{15\mathcal{G}m_2}{a_1^3}f_2(e_1)\Bigg)\\
\displaystyle Z_p &= \frac{m_1 k_p R_p^5}{\mu n a_1^5}\Bigg(\frac{2\OmBz^2-\OmBy^2-\OmBx^2}{2(1-e_1^2)^2}+\frac{15\mathcal{G}m_1}{a_1^3}f_2(e_1)\Bigg)
\end{align}
\end{subequations}

In terms of the previous definitions, the secularized non-conservative (nc) effects on the inner orbit and on the spin of the inner bodies are given by: 

\begin{subequations}
\label{eq:tidal_evol}
\begin{align}
\displaystyle &\bigg(\frac{de_1}{dt}\bigg)_{\mathrm{nc}} = -e_1(V_s+V_p)\\
\displaystyle &\bigg(\frac{da_1}{dt}\bigg)_{\mathrm{nc}} = -2a_1(W_s+W_p)\\
\displaystyle &\bigg(\frac{di_1}{dt}\bigg)_{\mathrm{nc}} = -\sin g_1 (Y_s+Y_p)+\cos g_1 (X_s+X_p)\\
\displaystyle &\bigg(\frac{dg_1}{dt}\bigg)_{\mathrm{nc}} = (Z_s+Z_p) - \cot i_1[\cos g_1 (Y_s+Y_p) + \sin g_1 (X_s+X_p)] \\
\displaystyle &\bigg(\frac{dh_1}{dt}\bigg)_{\mathrm{nc}} = \frac{\cos g_1}{\sin i_1}(Y_s+Y_p) + \frac{\sin g_1}{\sin i_1}(X_s+X_p)\\
\displaystyle &\bigg(\frac{d\Omega_{s,x}}{dt}\bigg)_{\mathrm{nc}} = \frac{m_r J_1}{I_s}\big[X_s(-\cos h_1 \sin g_1 - \sin h_1 \cos g_1 \cos i_1) + W_s \sin i_1 \sin h_1 - Y_s(\cos h_1 \cos g_1 - \sin h_1 \sin g_1 \cos i_1)\big]\\
\displaystyle &\bigg(\frac{d\Omega_{s,y}}{dt}\bigg)_{\mathrm{nc}} = \frac{m_r J_1}{I_s}\big[X_s(-\sin h_1 \sin g_1 + \cos h_1 \cos g_1 \cos i_1) - W_s \sin i_1 \sin h_1 - Y_s(\sin h_1 \cos g_1 + \cos h_1 \sin g_1 \cos i_1)\big]\\
\displaystyle &\bigg(\frac{d\Omega_{s,z}}{dt}\bigg)_{\mathrm{nc}} = \frac{m_r J_1}{I_s}\big[X_s\sin i_1 \cos g_1 + W_s \cos i_1 - Y_s \sin i_1 \sin g_1\big]\\
\displaystyle &\bigg(\frac{d\Omega_{p,x}}{dt}\bigg)_{\mathrm{nc}} = \frac{m_r J_1}{I_p}\big[X_p(-\cos h_1 \sin g_1 - \sin h_1 \cos g_1 \cos i_1) + W_p \sin i_1 \sin h_1 - Y_p(\cos h_1 \cos g_1 - \sin h_1 \sin g_1 \cos i_1)\big]\\
\displaystyle &\bigg(\frac{d\Omega_{p,y}}{dt}\bigg)_{\mathrm{nc}} = \frac{m_r J_1}{I_p}\big[X_p(-\sin h_1 \sin g_1 + \cos h_1 \cos g_1 \cos i_1) - W_p \sin i_1 \sin h_1 - Y_p(\sin h_1 \cos g_1 + \cos h_1 \sin g_1 \cos i_1)\big]\\
\displaystyle &\bigg(\frac{d\Omega_{p,z}}{dt}\bigg)_{\mathrm{nc}} = \frac{m_r J_1}{I_p}\big[X_p\sin i_1 \cos g_1 + W_p \cos i_1 - Y_p \sin i_1 \sin g_1\big]
\end{align}
\end{subequations}

Where the moment of inertia of the inner components are:

\begin{equation}
\displaystyle I_i = \kappa_s m_i R_i^2
\end{equation}

\noindent with $\kappa_i$ ($i=s,p$) are the so-called gyration radius of the involved bodies.


\bsp	
\label{lastpage}
\end{document}